 \documentclass[apj]{emulateapj}
 \usepackage{apjfonts}

\usepackage{amsmath}
  
\newcommand{\cmsq}{cm$^{-2}$}
\newcommand{\cmcc}{cm$^{-3}$}
\newcommand{\cps}{ct~s$^{-1}$}
\newcommand{\kms}{km~s$^{-1}$}

\newcommand{\ergps}{erg~s$^{-1}$}

\newcommand{\chan}{\textit{Chandra}}
\newcommand{\xmm}{\textit{XMM-Newton}}

\newcounter{subfigure}

\slugcomment{To appear in the Astrophysical Journal}

\shorttitle{\chan\  and \xmm\  Observations of YY Men}
\shortauthors{Audard et al.}

\begin{document}


\title{Some Like it Hot: The X-Ray Emission of The Giant Star YY Mensae}

\author{Marc Audard}
\affil{Columbia Astrophysics Laboratory, Mail code 5247, 550 West
120$^\mathrm{th}$ Street, New York, NY 10027}
\email{audard@astro.columbia.edu}

\author{Alessandra Telleschi, Manuel G\"udel}
\affil{Paul Scherrer Institut, W\"urenlingen \& Villigen, 5232 Villigen PSI, Switzerland}
\email{atellesc@astro.phys.ethz.ch, guedel@astro.phys.ethz.ch}

\author{Stephen L. Skinner}
\affil{Center for Astrophysics and Space Astronomy, University of Colorado, Boulder, CO 80309-0389}
\email{skinners@casa.colorado.edu}

\author{Roberto Pallavicini}
\affil{Osservatorio Astronomico di Palermo, Piazza del Parlamento 1, 90134
Palermo, Italy}
\email{pallavic@oapa.astropa.unipa.it}

\and

\author{Urmila Mitra-Kraev}
\affil{Mullard Space Science Laboratory, University College London, Holmbury St.
Mary, Dorking, Surrey RH5 6NT, United Kingdom}
\email{umk@mssl.ucl.ac.uk}

\begin{abstract}
We present an analysis of the X-ray emission of the rapidly rotating giant star
\objectname{YY Mensae}  observed by \chan\   HETGS
and \xmm. The high-resolution spectra display numerous emission lines of 
highly ionized species;  \ion{Fe}{17} to \ion{Fe}{25} lines are detected, together
with H-like and He-like transitions of lower $Z$ elements.
Although no obvious flare was detected, the X-ray luminosity changed by a factor of two 
between the \xmm\  and  \chan\  observations taken 4 months apart (from $\log L_X  \approx 32.2$ to $32.5$~\ergps,
respectively). The coronal abundances and the emission measure distribution have been derived from 
three different methods using optically thin collisional ionization equilibrium
models, which is justified by the absence of opacity effects in YY Men as
measured from line ratios of \ion{Fe}{17} transitions. The abundances show a distinct pattern as a 
function of the first ionization potential (FIP) suggestive of an inverse FIP effect as seen in
several active RS CVn binaries. The low-FIP elements ($<10$~eV)
are depleted relative to the high-FIP elements; when compared to its
 photospheric abundance, the coronal Fe abundance also appears depleted.
We find a high N abundance in YY Men's corona which we interpret as a signature
of material processed in the CNO cycle and dredged-up in the giant phase.
The corona is dominated by a very high temperature ($20-40$ MK) plasma, which places YY Men among the magnetically 
active stars with the hottest coronae.  Lower temperature plasma also
coexists, albeit with much lower emission measure.
Line broadening is reported in some lines, with a particularly strong significance in \ion{Ne}{10} Ly$\alpha$.
We interpret such a broadening as Doppler thermal broadening, although rotational broadening due to 
X-ray emitting material high above the surface could be present as well. We use two different formalisms
to discuss the shape of the emission measure distribution. The first one infers the
properties of coronal loops, whereas the second formalism uses flares as a statistical ensemble.
We find that most of the loops in the corona of YY Men have their maximum temperature equal to
or slightly larger than about $30$~MK. We also find that 
small flares could contribute significantly to the coronal heating in YY Men.
Although there is no evidence of flare variability in the X-ray light curves,
we argue that YY Men's distance and X-ray brightness does not allow us to
detect flares with peak luminosities $L_\mathrm{X} \leq 10^{31}$~\ergps\  
with current detectors. 
\end{abstract}


\section{Introduction}

Magnetic activity is ubiquitous in late-type stars, although its level can vary
dramatically. A common activity indicator is the ratio of the X-ray luminosity to
the stellar bolometric luminosity, $L_\mathrm{X} / L_\mathrm{bol}$; it
varies typically from $10^{-7}$ in inactive stars to $10^{-3}-10^{-2}$ in the
most active stars (see, e.g., \citealt{favata03} and \citealt{guedel04} for recent
reviews on the X-ray emission of stellar coronae). Giant stars present
a peculiar behavior in which magnetic activity is strong for spectral types 
earlier than typically K0-2 III, while it vanishes rapidly for cooler spectral 
types \citep[e.g.,][and references therein]{linsky79,ayres81,linsky85,huensch96b}.
Rotational breaking occurs in giants around spectral type G0~III \citep{gray89}. 
Although the origin of the divide remains debated, \citet{huensch96a} suggested that 
the evolutionary history of
magnetically active stars of different masses explains the X-ray dividing 
line \citep*[also][]{schroeder98}. Alternatively, \citet{rosner95} proposed 
that the dividing line could be explained by the change from a predominantly 
closed magnetic configuration in stars blueward of the dividing line to a 
predominantly open configuration in cooler giants. The latter 
configuration induces cool winds and the absence of hot coronae. 
\citet{holzwarth01} argued that flux tubes in giants of spectral 
types G7 to K0 remain inside the stellar convection zone in a stable 
equilibrium.

FK Comae stars form a loosely-defined class of rapidly rotating single
G and K giant stars, whose outstanding property is a projected
equatorial velocity measured up to 160~\kms, in contrast to the
expected maximum of 6~\kms\  for giants. One of the leading theories to
explain the extreme properties of FK Com stars suggests that they were
formed by coalescence of a contact binary when one of the components entered
into the giant stage \citep{bopp81a,bopp81b,rucinski90}. \citet{simon89} noted
alternatively that dredge-up of angular momentum during the early red giant
phase could explain their rapid rotation. Magnetic activity in FK
Com stars is very strong \citep{bopp81b,fekel86,rutten87,simon89}. In the X-ray regime, their 
X-ray luminosities are in the range from $L_\mathrm{X} = 10^{30}$ to $10^{31}$~\ergps,
i.e., $L_\mathrm{X} / L_\mathrm{bol} \sim 10^{-5} - 10^{-3}$ 
\citep*[e.g.,][]{maggio90,welty94,demeideros95,huenemoerder96,huensch98,gondoin99,gondoin02}.

We present in this paper new high-resolution X-ray spectra and high
signal-to-noise light curves of the FK Com-type star \objectname{YY
Mensae} (\objectname{HD 32918}) obtained with the \textit{Chandra} X-ray
Observatory and its High-Energy Transmission Grating Spectrometer (HETGS), and with
the \textit{XMM-Newton} Observatory. With these observations, we aimed
to compare the X-ray emission of this bright FK Com-type star with that
of other magnetically active stars. We show that YY Men
is among the stars with the hottest coronae, with a dominant plasma temperature
around $20-40$~MK. Furthermore, we investigated the elemental composition of YY Men's 
corona to study abundance anomalies. 

The paper is structured as follows: Section \ref{sect:yymen} describes the 
main properties of YY Men and previous
observations; section \ref{sect:obs} gives the observation details, whereas
the data reduction is described in section \ref{sect:reduction}
(Sect.~\ref{sect:chandra_data} for \chan\  and Sect.~\ref{sect:xmm_data} for \xmm). In
section \ref{sect:lc}, we provide an analysis of the light curves. Section
\ref{sect:spectral} describes the procedures for our spectral analysis for 
\chan\  (Sect.~\ref{sect:chandra_spectral}) and \xmm\  (Sect.~\ref{sect:xmm_spectral}).
We approached the spectral inversion problem with three different methods to obtain the emission
measure distribution (EMD) and abundances in YY Men's corona
(Sect.~\ref{sect:methodology}). Section
\ref{sect:discussion} includes a discussion of our results: abundances (Sect.~\ref{sect:abundances}), EMD
(Sect.~\ref{sect:emd}), line broadening
(Sect.~\ref{sect:doppler}), electron densities (Sect.~\ref{sect:densities}), and optical
depth effects (Sect.~\ref{sect:opacities}). Finally, we give a summary and our
conclusions in section \ref{sect:conclusions}.

\section{The Giant Star YY Mensae}
\label{sect:yymen}

\objectname{YY Mensae} (K1~IIIp) is a strong \ion{Ca}{2} emitter \citep{bidelman73} at 
a distance of $291$~pc \citep{perryman97}, with a mass of $M = 2.2~M_\sun$ \citep[see][]{gondoin99}, and
a radius of $R = 8.8 \times 10^{11}$~cm = $12.7 R_\sun$, 
based on $T_\mathrm{eff} = 4,700$~K from \citet*{randich93} and $L_\mathrm{bol} = 2.7 \times 10^{35}$~\ergps\  
$ = 70 L_\sun$ from \citet{cutispoto92}. It belongs to the FK Com-type class of giant
stars \citep{collier82}.
It displays a photometric period of $P = 9.55$~days, a projected rotational
velocity of $v \sin i = 45 - 50$~\kms\  \citep*{collier82,piskunov90,glebocki00}, and
an inclination angle of $i = 65\arcdeg$ \citep{piskunov90}.
\citet*{grewing86} presented the \textit{IUE} ultraviolet spectrum of
\objectname{YY Men}, showing very strong ultraviolet (UV) emission lines originating 
from the chromosphere and the transition zone. They obtained YY Men's EMD 
in the range $\log T (\mathrm{K}) = 3.8 - 5.3$, emphasizing its
strength in comparison to other magnetically active stars. Line ratios
indicated characteristic electron densities in the transition zone of $n_e = 3 -
4 \times 10^{10}$~\cmcc\  \citep{grewing86}. Strong flares were
observed in the radio and optical, lasting for several days 
\citep*{slee87,bunton89,cutispoto92}. A strong P Cyg profile in the H$\alpha$
line during a flare indicated strong outflowing material with a
velocity of $230$~\kms\  \citep{bunton89}. Finally, spot coverage in YY Men was 
generally found in an equatorial belt \citep{piskunov90}, in
contrast to the large polar caps found in many active stars
\citep[e.g.,][]{vogt88}. 

In the X-ray regime, \citet*{bedford85} reported X-rays from YY Men detected
with the \textit{EXOSAT} CMA instrument. \citet{guedel96} gave results of
observations made with \textit{ROSAT} and \textit{ASCA}. They found indications
for a hot (up to 3~keV) dominant coronal plasma. Flaring activity was reported,
although hampered by many interruptions due to the low orbits of the satellites.
Abundance depletion in most elements was suggested from fits to the medium-resolution 
\textit{ASCA} spectra. The X-ray luminosities varied between the observations
and ranged from $\log L_\mathrm{X} = 32.3$ to $32.65$~\ergps\  \citep{guedel96}.

Our \textit{Chandra} and \textit{XMM-Newton} observations are generally
consistent with the above results. Nevertheless, deep, uninterrupted 
monitoring and high spectral resolution provide the opportunity to significantly improve
our understanding of the corona of YY Men.

\section{Observations}
\label{sect:obs}
\chan\  and \xmm\  observed YY Men as part of Cycle 4 and of the
Reflection Grating Spectrometer (RGS) Guaranteed Time, respectively. We provide
a log of the observations in Table~\ref{tab:log}.

\chan\  observed the giant with the Advanced CCD Imaging Spectrometer (ACIS) in
its 1x6 array (ACIS-S) with the HETG inserted \citep{weisskopf02}. 
This configuration provides 
high-resolution X-ray spectra from 1.2 to 16~\AA\  for the High-Energy Grating 
(HEG) spectrum and from 2.5 to 31~\AA\  for the Medium-Energy Grating (MEG)
spectrum. The HETGS has constant resolution (HEG: $\Delta\lambda \sim 12$~m\AA, MEG:
$\Delta\lambda \sim 23$~m\AA\  FWHM).
Thanks to the energy resolution of the ACIS camera, spectral orders can be
separated. However, because of the drop in effective area with higher spectral orders, we used
only  the first-order spectra. Further details on the instruments can be found in the
\chan\ \anchor{http://cxc.harvard.edu/proposer/POG}{Proposers'
Observatory Guide}\footnote{\url{http://cxc.harvard.edu/proposer/POG}.}.

\xmm\  \citep{jansen01} observed YY Men in the X-rays with the RGS
 ($\lambda \sim 6-38$~\AA\  with $\Delta\lambda \sim
 60-76$~m\AA; \citealt{denherder01}) and the 
European Photon Imaging Camera (EPIC) MOS only ($0.15-10$~keV with
$E/\Delta E = 20-50$; \citealt{turner01}), since the EPIC
pn camera \citep{strueder01} was off-line due to a sudden switching off of one 
quadrant \citetext{M. Guainazzi, priv. comm.}. Simultaneous optical coverage was 
obtained with the Optical Monitor \citep[OM;][]{mason01}, although we did not
make use of the data because YY Men was too bright ($V=8.1$) for
the $U\!$ filter. Furthermore, we note that, due to a small pointing misalignment,
YY Men fell off the small $11\arcsec \times 11\farcs 5$ timing window.

\section{Data Reduction}
\label{sect:reduction}

\subsection{The \chan\  data}
\label{sect:chandra_data}The \chan\  data 
were reduced with the \chan\  Interactive Analysis of Observations
(CIAO) version 2.3 in conjunction with the calibration database (CALDB) 2.21.
We started from the level 1 event file. We used standard techniques described in
analysis threads\footnote{\url{http://cxc.harvard.edu/ciao/threads}.}. In
particular, we applied a correction for charge transfer inefficiency,
applied pulse height analysis and pixel randomizations, and we destreaked CCD 8 (ACIS-S4).
Furthermore, we used non-default masks and background spectral extraction masks
to reflect the modifications introduced in the recent CIAO 3 release\footnote{
We used \texttt{width\_factor\_hetg=35} with task \texttt{tg\_create\_mask},
and latter used \texttt{max\_upbkg\_tg\_d=6.0e-3} with task
\texttt{tgextract}.}.

We then calculated grating response matrix files (RMFs) and grating ancillary response
files (ARFs). CALDB 2.21 contained separate line spread functions (LSF) for 
the positive and the negative order spectra. We then co-added the positive and
negative order spectra to increase statistics (using the co-added grating ARFs), and further binned the spectra by
a factor of two. Our final HEG and MEG first-order spectra have bin widths of
2.5 and 5~m\AA, respectively. The effective exposure was 74.2~ks. Finally, we note 
that no contaminating X-ray source was detected in the zeroth order image.

\subsection{The \xmm\  data}
\label{sect:xmm_data}
The \xmm\  X-ray data 
were reduced with the Science Analysis System (SAS)
version 5.4.1 using calibration files from August 2003. The RGS data were reduced
with \texttt{rgsproc}. The source spatial mask included 95\% of the cross-dispersion
function, whereas the background spatial mask was taken above and below the
source, by excluding 97\% of the cross-dispersion function. The mask in the
dispersion-CCD energy space selected first-order events only and included 95\%
of the pulse-invariant energy distribution. We generated RGS RMFs with 6,000
energy bins. The maximum effective exposure for the RGS was about 85~ks. 

Among the EPIC data, we used the EPIC MOS1 data only since we preferred to give maximum weight to 
the high-resolution RGS spectra. We removed short periods of solar flare 
activity (4.25~ks), leaving $\sim 81$~ks of MOS1 exposure time.
We extracted the MOS1 data from a circle (radius of 47\arcsec) around the source, and the
background from a source-free region of the same size on a nearby outer CCD.
We corrected the exposure for vignetting in the background spectrum.
We finally created MOS1 RMF and ARF  files for the source, using default parameters in SAS
5.4.1 (except for the detector map type, for which we used the observed
instrumental PSF at 3~keV).

\section{Light curves}
\label{sect:lc}

We extracted light curves from the \chan\  and the \xmm\  observations. Since the
zeroth order \chan\  light curve was piled up, we used the dispersed
first-order photons instead. The background for the dispersed photons was taken
from two rectangles ``above'' and ''below'' the source in the
dispersion/cross-dispersion space, totaling an area eight times larger than the
source. The scaled \chan\  background was, however, about 150 times fainter 
than the source. The EPIC MOS1 light curve is based on the extraction regions
described in Sect.~\ref{sect:xmm_data}. Note that the light curves are not corrected
for deadtime which is in any case negligible.

\begin{figure}
\centering
\includegraphics[width=\textwidth]{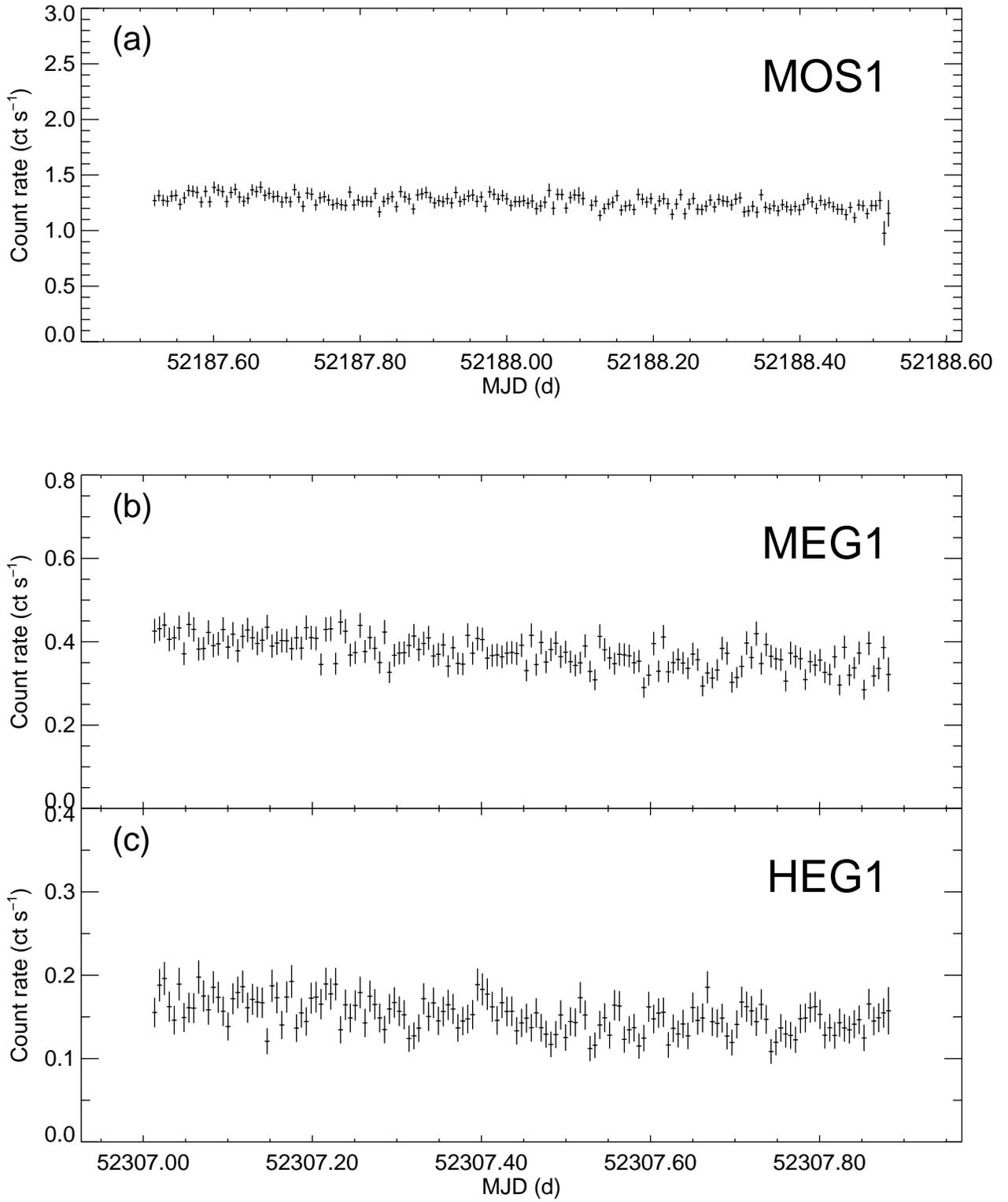}
\caption{Background-subtracted light curves of YY Mensae from \xmm\  MOS1 (a),
\chan\  MEG1 (b), and HEG1 (c) for a bin size of 500~s.\label{fig:lc}}
\end{figure}

Figure~\ref{fig:lc} shows the \xmm\  MOS1, and the
\chan\  MEG and HEG first-order light curves with a time bin size of 500~s. 
YY Men was about $2$ times brighter ($\log L_\mathrm{X} \sim 32.5$~\ergps) during 
the \chan\ observation  than during the \xmm\  observation ($\log L_\mathrm{X} \sim 32.2$~\ergps), 
as derived from fits to the average spectra (Sect.~\ref{sect:spectral}).
Both the \xmm\  and \chan\  light curves show no obvious flare, but they display
a slow decrease in flux with time. The modulation is weak ($15-20$\%), but
a Kolmogorov-Smirnov test for the MOS data gives a very low probability of constant count 
rate ($P \ll 0.1$). We also performed a Kolmogorov-Smirnov test to search for short-term 
variability in the \xmm\  MOS light curve. None was found down to a time scale 
of $\sim 300$~s.

The interpretation of the modulation is unclear. We believe that
YY Men was not caught in the late phase of a flare decay,
since there is no evidence of a temperature variation during 
the observations. However, it could be due
to some rotational modulation effect in which active regions
almost completely cover the surface of the very active YY Men.

\section{Spectral Analysis}
\label{sect:spectral}

In this section we present the spectral analysis of YY Men's data taken with \chan\  
and \xmm. Due to the dominant hot corona  and to the high derived
interstellar absorption ($N_\mathrm{H} \sim 7 \times 10^{20}$~\cmsq; Tab.~\ref{tab:meth}; see below as well) 
toward YY Men, the \chan\  HETGS spectrum is 
the best-suited grating instrument; the \xmm\  data were, however, most useful to access
the long wavelength range (and thus the C abundance).

\subsection{\chan}
\label{sect:chandra_spectral}
We fitted the co-added first-order spectra of MEG and HEG simultaneously.
However, we restricted the wavelength range to $1.6 - 16.2$~\AA, and 
$1.8 - 25.0$~\AA\  for HEG and MEG, respectively. 
Thanks to the low ACIS background during the observation it was not necessary to
subtract  a background spectrum. Each background spectral bin contained 
typically $0-2$ counts in HEG and $0-4$ in MEG \emph{for an area 8 times larger
than the source}, thus the scaled contribution of the background is less than
a half (a quarter) of a count per bin in MEG (HEG), much less than in the MEG 
or HEG first-order source spectra of YY Men. It allowed us to use the robust $C$ statistics 
\citep{cash79}. 

\subsection{\xmm}
\label{sect:xmm_spectral}
The \textit{XMM-Newton} RGS1, RGS2, and MOS1 data were fitted simultaneously\footnote{Although
cross-calibration effects could, in principle, be of some importance, we verified that it was
not the case for YY Men by obtaining multi-$T$ fits to the MOS1 data alone, and to the RGS1+RGS2
spectra as well. The best-fit solution to the MOS1 spectrum is close to the solution reported
in Table~\ref{tab:meth} for the combined spectra. On the other hand, due to the high $T$ of YY Men, the RGS data
alone were not sufficient to constrain adequately the high-$T$ component, and consequently absolute
abundances are different. Nevertheless, abundance ratios relative to Fe
are similar to the ratios found for the combined fit. We therefore feel confident that
the combined RGS+MOS fits are not strongly biased by cross-calibration effects.}.
The spectra were grouped to contain a minimum of 25 counts per bin.
We used the RGS spectra longward of $8.3$~\AA, and 
discarded the EPIC data longward of 15~\AA\  due to the lack of spectral resolution of the CCD spectrum;
in addition, we preferred to put more weight on the high-resolution RGS data.
As the \xmm\  data are background-subtracted, we used the $\chi^2$ statistics.

\subsection{Methodology}
\label{sect:methodology}

We approached the spectral analysis with three different methods to obtain the EMD and
abundances in YY Men's corona. We aim to compare the
different outputs in order to discuss the robustness of the results. For each
method, the coronal abundances in YY Men were compared to the solar photospheric abundances 
given in \citet{grevesse98}.
The first method uses the classical
multi-$T$ model to discretize the EMD. 
The second method obtains a continuous EMD described
by Chebychev polynomials. 
The third method uses selected emission line fluxes to derive a
continuous EMD and abundances. The continuum contribution is estimated from
line-free continuum spectral bins. We describe the above methods in 
the following sections in detail.

\subsubsection{Method One: A Multi-Temperature Discretization of the Emission Measure Distribution}

This method uses the classical approach of a multi-$T$ component
model as a discretization of the EMD. The model, however, does not provide
a physically reasonable description of the corona of a star since EMDs are thought to 
be continuous. Nevertheless, this approach is generally sufficient to obtain 
abundances \citep[e.g.,][]{schmitt04}.

We applied method 1 to the \chan\  and \xmm\  data. We used 
the Interactive Spectral Interpretation System
(ISIS) software version 1.1.3 \citep{houck00} for \chan\  and XSPEC \citep{arnaud96}
for \xmm. The collisional ionization  equilibrium models were based on
the Astrophysical  Plasma Emission Code (APEC) version 1.3 \citep{smith01},
which was available in both software packages. 
Coronal abundances were left free, in addition to the $T$ and EM.
We included a photoelectric absorption component left free to vary that uses cross-sections from
\citet{morrison83}. We included thermal broadening for the \chan\  spectra. 
While such a broadening is negligible for most lines, it plays some
role for certain ions (see Sect.~\ref{sect:doppler}).

\notetoeditor{The following figure is divided into three sub-figures which 
should be labeled as (a), (b), and (c).}

Figures~\ref{fig:chandra_bfit_meth1a}--\ref{fig:chandra_bfit_meth1c} show the \textit{Chandra} HEG and MEG spectra
with the best-fit 3-$T$ model overlaid. The overall agreement is very
good. Figure~\ref{fig:xmmspec} shows the \xmm\  spectra with the best-fit
4-$T$ model overlaid. The higher sensitivity of \textit{XMM-Newton} RGS at long
wavelengths over \textit{Chandra} HETGS allowed us to detect
a low-$T$ ($\sim 3$~MK) plasma component which was not detected with
\textit{Chandra}. Although the component is faint (as expected from
the \textit{Chandra} data), its inclusion decreases the $\chi^2$ statistics significantly
($\Delta\chi^2 = 32$ for $2$ additional degrees of freedom).

\subsubsection{Method Two: A Continuous Emission Measure Distribution from Chebychev Polynomials}

This method obtains a continuous EMD described by Chebychev polynomials. 
An additional constraint for the spectral inversion
problem is to keep EMs positive. This was achieved by approximating the EMD
with the exponential of a polynomial as described by \citet{lemen89}.
We use the convention that the differential emission measure, $\varphi(T)$, 
is defined as
\begin{equation}
\varphi (T) = n_\mathrm{H} n_\mathrm{e} \frac{dV}{dT} \qquad\text{(\cmcc~K$^{-1}$)}.
\label{eq:dem}
\end{equation}
Thus the total EM is given by $\mathrm{EM_{tot}} = \int \varphi (T) dT = \int \varphi(T) T d(\ln T)$.
A graphical representation of the EMD is therefore given by
\begin{equation}
\mathrm{EMD}(T) = \varphi(T) T \Delta\log T \qquad\text{(\cmcc)}.
\label{eq:emd}
\end{equation}
In this paper, we preferred to plot $\mathrm{EMD}(T) / \Delta\log T$ instead, 
since it is independent of the grid bin size ($\Delta\log T$), to help
comparison between the EMD reconstructed with methods 2 and 3.
We constructed a local model in XSPEC \citep{arnaud96} in which the maximum polynomial degree
can be fixed, and in which the $T$ grid can be given as well\footnote{XSPEC has similar models, e.g., \textsc{c6pvmkl},
however, the models use the MEKAL database, a fixed $T$ grid from $\log T = 5.5$
to $8.0$ with $\Delta\log T = 0.10$, and a sixth-order polynomial only.}. 
Coronal abundances and photoelectric absorption were also left free to vary.
Our model uses the same APEC version as in ISIS. We applied method 2 to both 
the \chan\  and \xmm\  spectra. 

We report here the results obtained with a grid between $\log T = 6.0$ and $7.95$~K and a bin width
of $\Delta\log T =0.15$~dex, 
and for a polynomial degree of $n = 7$. Various combinations of $T$ ranges, grids,
and polynomial degrees were tested as well but they provided no improvement of the statistics.
The $T$ grid is fixed, and EMs are obtained from Chebychev coefficients $a_k$ ($k = 1,\dots,7$) and
from a normalization factor in XSPEC.
We determined uncertainties on EMs for each $T$ bin
as follows: we obtained 68\% ($\Delta C$ or $\Delta\chi^2  = 1$) confidence ranges for the Chebychev
polynomials $a_1,\dots,a_7$, and assumed that each coefficient was normally distributed
with a mean, $\mu_i$, equal to the best-fit value and with a standard deviation, $\sigma_i$, equal to the
geometric mean of its 68\% uncertainties. We then used a Monte-Carlo approach and, for each coefficient,
we generated $10,000$ pseudo-random values following a normal distribution 
$\mathcal{N}(\mu_i,\sigma_i)$  (since Chebychev coefficients are in the range $[-1,1]$, 
we assigned a value of $-1$ if the randomly generated value was $< -1$, and assigned $+1$ if the random
number was $>1$). Our Monte-Carlo approach provided us
with 10,000 values of the EM per $T$ bin. We investigated the distribution of
each EM per bin and noted that its logarithmic value closely followed a normal
distribution. We therefore fitted the 13 distributions with normal functions, and associated
the derived standard deviation with the uncertainty of the EM per bin. 

\subsubsection{Method Three: A Continuous Emission Measure Distribution 
Derived from Fe Emission Lines}\label{sect:method3}

The last method used extracted fluxes from specific, bright lines to reconstruct the
EMD. We applied it only to the \chan\  data given the low line-to-continuum ratios and the 
broad wings in the \xmm\  RGS spectrum, both of which make a clean flux extraction
challenging in this case.

To extract line fluxes, we used the \chan\   HEG1 and MEG1 data simultaneously to fit each emission line
together with a ``line-free'' continuum in ISIS. 
The continuum model consisted of a single $T$, absorbed model
containing the two-photon emission, radiative recombination, 
and bremsstrahlung continua and a pseudo-continuum. The pseudo-continuum in APEC consists of lines 
which are too weak to list individually and whose contributions are stored as a continuum.
The fitting procedure consistently
found a best-fit $\log T = 7.53$~K with $\log {\rm EM} = 55.25 \pm 0.02$~\cmcc\  and an absorption
column density of $\log N_\mathrm{H} = 20.89 \pm 0.02$~\cmsq.

Next, we applied an iterative EMD reconstruction method using the above line fluxes and the
same APEC database as with the other methods. 
To allow for comparison with method 2, we used $\mathrm{EMD}(T) / \Delta\log T$ for our graphical representation
of the EMD.
The method is described in detail in \citet{telleschi04}; however, we briefly 
summarize the principal steps. We treated unresolved line blends containing essentially
only lines of one element like single lines, i.e., we computed new
$T$-dependent emissivities for each considered line blend.
Our EMD reconstruction starts by considering the Fe lines of Fe line blends. An
approximate, smooth EMD estimated from the emissivities at the maximum line formation 
$T$ for each line serves as an initial approximation to the
solution. The coolest portion of the EMD is estimated from the flux ratio of the 
\ion{O}{8} to  the \ion{O}{7} resonance lines. The EMD is binned into bins of width 
$\Delta\log T = 0.1$~dex in the $T$ range between $\log T (\mathrm{K})= 6-8$. 
Once an EMD is defined, line fluxes were predicted by integrating the emissivities 
across the EMD. We then iteratively corrected the 
EM in each bin, using the iteration algorithm described by \citet{withbroe75}.
The iteration is terminated once the $\chi^2$ (for the deviations
between predicted and measured fluxes) is no longer significantly
improving, or if the reduced $\chi^2$ is $\le 1$. 
At this point, the EMD has been determined only up to a normalization factor that directly depends on 
the absolute [Fe/H] abundance. To  determine these two quantities, we computed the predicted spectra for 
various values of [Fe/H] until the predicted continuum was in close agreement with the observations.  
We repeated this entire analysis twenty times, each time perturbing the measured line fluxes according 
to their measurement errors and an assumed systematic uncertainty in the line emissivities (10\% for 
each line). The scatter in the solutions provided an estimate of the uncertainties in the EMD.  
Abundance uncertainties include
i)  the measurement errors of the fluxes of the line blends, 
ii) the scatter in the abundance values derived from the different lines of the same element, 
iii) the scatter in the abundance values derived from the twenty EMD realizations, 
and, iv) for the absolute Fe abundance,  
 the statistical uncertainty in the adjustment to the observed 
continuum level.

\subsubsection{Systematic Uncertainties}

\notetoeditor{Place Table \ref{tab:meth} with this section}

Table~\ref{tab:meth} shows the best-fit parameters for \chan\  and \xmm\  for the 3 methods
discussed above. We provide for each fit parameter an estimate of the uncertainty based on the 
confidence  ranges for a single parameter of interest (90~\%, except for method 3 where
we quote 1-$\sigma$ ranges). However, we emphasize that such uncertainties are purely
statistical, and do not include systematic uncertainties except for method 3, see \S\ref{sect:method3}. 
Those are typically of
instrumental nature (e.g., cross-calibration of the MEG and HEG, wavelength
scale, effective area). In addition, models using atomic data do not usually
include uncertainties for the atomic parameters (e.g., transition wavelengths,
collisional rates, etc). These uncertainties (of the order
of 10-50\%; \citealt{laming02}) vary from element to element and from transition to 
transition, and thus are difficult to estimate as a whole. Certain groups of
atomic transitions, like low-Z L-shell transitions, are also simply missing 
in many atomic databases \citep[e.g.,][]{lepson03}. If the temperature structure
is such that those transitions are bright enough to be measured in
an X-ray spectrum, their absence in the atomic database can have a significant
impact on the determination of the EMD and of abundances. Typically those lines
are maximally formed at cool temperatures and can be found longward of 25~\AA\ 
\citep[e.g.,][]{audard01a},
thus they are weak for most hot coronal sources (an example where weak lines are important
is the cool F subgiant Procyon; \citealt{raassen02}). The \chan\  and
\xmm\   X-ray spectra of YY Men's hot corona are, therefore, not strongly
contaminated by these L-shell lines, which allowed us to include the relevant
wavelength ranges. A small contamination could potentially still
contribute to the systematic uncertainty of coronal abundances.
In addition, since the spectral inversion problem to construct
an EMD is mathematically ill-posed, EMDs are not unique, and several 
realizations can reproduce the observed spectrum \citep{craig76a,craig76b}.
Finally, uncertainties in the solar photospheric abundances exist
as well. All the above contribute to systematic uncertainties. 
Consequently a systematic uncertainty of at least $0.1$~dex in the 
EMs and abundances should be expected.

\section{Discussion}
\label{sect:discussion}

In this section, we discuss quantities derived from the spectral and photometric
analyses with \textit{Chandra} and \textit{XMM-Newton}. We provide comparisons
between the observations obtained at two different epochs when relevant.

\subsection{Abundances}
\label{sect:abundances}
Our analyses of the \chan\  HETGS and \xmm\  spectra 
showed a marked depletion of metals in YY Men's corona when compared to the
solar photospheric composition \citep{grevesse98}. There is generally good agreement between the
abundances obtained with the various methods. Below we compare the abundances
obtained in this paper and discuss them in the context of an abundance pattern
in YY Men's corona.

\subsubsection{Systematic Effects}
\label{sect:abun_syst}

Various methods applied to the same data set gave relatively similar abundances within the error bars;
however, small systematic offsets can be observed. For example, abundances with
\chan\  data from method 2 are slightly lower with respect to 
those obtained from method 1 (Fig.~\ref{fig:ab_m12_chan123}a). This is
probably a trade-off of the fitting procedure between the EMD
discretization and the values of the abundances. On the other hand, 
no significant shift is visible in the \xmm\  data
(Fig.~\ref{fig:ab_m12_chan123}b). 
Such a comparison is useful, since it shows that despite using the same
data set, the energy range, and atomic database, systematic
offsets in abundances can be derived from spectral fits \citep{craig76b}. We emphasize,
however, that such an offset is small in the case of YY Men where we find excellent
agreement.

Method 3 was applied to the \chan\  data only, and by definition, abundance
ratios with respect to Fe are obtained, and later the absolute Fe abundance is determined from the continuum.
We compare the abundance ratios obtained with methods 1 and 3 in Figure~\ref{fig:ab_m12_chan123}c
and with methods 2 and 3 in Figure~\ref{fig:ab_m12_chan123}d.
\emph{Excellent agreement is derived for most elements,
demonstrating that abundance ratios are very stable regardless of the method used}. We find
no preference for either (so-called) global fits or line-based analysis. 
We note, nevertheless, weaker agreement for the Ca abundance
ratio, which is larger with method 3 than with both methods 1 and 2. The
reported abundances with method 3 are weighted averages determined from the
available lines of an element, as derived from the line fluxes and the shape of
the EMD. Ca was, however, determined
from \ion{Ca}{19} only because \ion{Ca}{20} is not obvious in the spectra.
On the other hand, the first two methods could underestimate the Ca abundance.
Indeed, the \ion{Ca}{19} triplet is weaker in the model than in the MEG spectrum
(Fig.~\ref{fig:chandra_bfit_meth1a}). The robustness of abundance ratios, even
with the simplistic method 1, is an important result that can be explained by
the fact that our models (including the multi-$T$ approach) sampled the EMD adequately to
cover the cooling function of each line.

We also compared the abundances obtained with the same method and atomic
database, but with different data sets (Fig.~\ref{fig:ab_xmmchan}). The top
panels in Figure~\ref{fig:ab_xmmchan} show the absolute abundances (relative to
H) from the \chan\  data on the $x$-axis and from the \xmm\  data on the
$y$-axis for the first method (left) and the second method (right). 
The \xmm\  absolute abundances are generally lower by $\sim 0.15$~dex with 
respect to the \chan\  abundances, regardless of the method. In contrast, we find
a much better agreement when abundance ratios (e.g., with respect to Fe; bottom panels) are
used. It again indicates that such ratios are significantly more stable. 

Although the absolute N abundance is larger with \chan\  than with \xmm,
we find that the N/Fe abundance ratios match well within the uncertainties. 
Nevertheless, with \chan\  data, method 3 found a closer agreement to the \xmm\ 
ratio than methods 1 and 2, for
which the best-fit ratios are slightly larger. We attribute this effect to the
faintness of the \ion{O}{7} triplet in the \chan\  data, and the lack of coverage
at long wavelengths. Both methods 1 and 2 could not constrain accurately the
EMD at low temperature (where \ion{N}{7} is maximally formed), and thus the N abundance.
This is, indeed, reflected in the lack of cool ($\sim 3$~MK) plasma in the spectral fit to 
\chan\  with method 1. This component was required with the \xmm\  data and thus the fit 
needed a lower N abundance to describe the \ion{N}{7} Ly$\alpha$ line. Since the algorithm
for method 3 includes the ratio of the \ion{O}{8} line to the \ion{O}{7} line, there
is a better coverage of the EMD at low temperature.

The abundances of Ar and S 
are lower with \xmm\  than with \chan. Since lines are found mostly from
K-shell transitions in the EPIC CCD spectrum, possible cross-calibration errors
between the RGS and the EPIC could explain the discrepancy. We note, however,
that the upper limit of the Ca abundance with \xmm\  is consistent with the
abundance obtained with \chan. 
We note that an alternative solution could be that true abundance variations occurred 
in such elements between the \xmm\  and \chan\  observations. Although we cannot
strictly discard such an explanation, we find it improbable.

\subsubsection{An Inverse FIP Effect}
\label{sect:fip}

In contrast to the solar First Ionization Potential (FIP) effect in which
low-FIP ($<10$~eV) elements are overabundant with respect to the solar
photospheric composition and in which high-FIP elements are of photospheric
composition \citep*[e.g.,][]{feldman92,laming95,feldman00}, recent analyses of grating
spectroscopic data with \xmm\  and \chan\  have emphasized the presence of an
inverse distribution in very active stars
\citep*[e.g.,][]{brinkman01,drake01,audard01b,guedel01,huenemoerder01,audard03} in
which the low-FIP elements are depleted with respect to the high-FIP elements.
Unknown or uncertain photospheric abundances are, however, a general problem 
in magnetically active stars. Possibly, coronal abundances relative to the
\emph{stellar} photospheric abundances could show different patterns
\citep*[e.g.,][]{audard03,sanz04}. However, a study of solar analogs with
solar photospheric composition, and of various ages and activity levels showed
a transition from a solar-like FIP effect to an inverse FIP effect with
increasing activity \citep{guedel02}. Bright, active RS CVn binaries appear to
follow the transition generally well \citep{audard03}.
Several studies have focused on patterns of abundances in magnetically active 
stars \citep[e.g.,][]{drake96}. We refer the reader to recent reviews for further information 
\citep{drake03a,favata03,guedel04}.

Figure~\ref{fig:ab_fip} plots the abundance ratios with respect to Fe in YY Men's corona,
relative to the solar photospheric composition \citep{grevesse98}. As shown
previously, abundance ratios are more robust than absolute abundances in
spectral fits. The \xmm\  data also give access to the C abundance, whereas the
\chan\  data provide an estimate of the Al abundance. Three features can
be observed immediately in Fig.~\ref{fig:ab_fip}: i) low-FIP elements
($<10-15$~eV) show solar photospheric ratios, ii) high-FIP elements (essentially Ar and Ne,
and possibly O) are overabundant with respect to solar ratios, iii) N is highly
overabundant; this feature is addressed in Sect.~\ref{sect:cnocycle}.
The first two features are reminiscent of the inverse FIP effect
seen in many stars with high activity levels \citep[e.g.,][]{audard03}, such
as YY Men ($L_\mathrm{X} / L_\mathrm{bol}
\sim 10^{-3}$ using $L_\mathrm{bol} = 2.7 \times 10^{35}$~\ergps\  from 
\citealt{cutispoto92}). The low-FIP elemental abundance ratios are all consistent
with a solar ratio. In contrast, several active stars show a broad U-shape
in their FIP pattern, with Al and Ca slightly overabundant with respect to
Fe, whereas the abundances increase gradually with increasing FIP
\citep[e.g.,][]{huenemoerder03,osten03}.

Many studies of stellar abundances focused mainly on the [Fe/H] photospheric abundance
\citep*[e.g.,][]{pallavicini92,randich93,cayrel97,cayrel01}. Consequently, 
coronal FIP patterns can often only be compared with the solar photospheric
composition, except in (rare) cases where photospheric abundances of some
elements are relatively well-known \citep*[e.g.,][]{drake95,drake97,guedel02,raassen02,audard03,sanz04,telleschi04}. 
YY Men is no exception. To our knowledge, only the [Fe/H] abundance is available in the
literature. \citet{randich93} quote $\mathrm{[Fe/H] = -0.5}$; however, in their
spectrum synthesis analysis, they used a solar abundance of $\mathrm{A(Fe)=7.63}$, 
\citetext{S.~Randich 2003, priv. comm.}, while we use in this paper the
\citet{grevesse98} abundance, $\mathrm{A(Fe)=7.50}$. Therefore, we will use
$\mathrm{[Fe/H] = -0.37}$ in YY Men's photosphere. Together with the absolute
abundances, [Fe/H] $\sim$ $-0.5$ to $-0.65$, in YY Men's corona from Table~\ref{tab:meth}, we conclude that
YY Men's corona is depleted in Fe with respect to its stellar photosphere by
about $-0.15$ to $-0.30$~dex. Assuming that the abundances of the other elements 
are depleted by a similar amount ($-0.37$~dex) in YY Men's photosphere, we
cautiously suggest that low-FIP elements are depleted in its corona relative to its
photosphere as well, 
whereas high-FIP elements are either photospheric or slightly enhanced.
Obviously, additional photospheric data are needed to substantiate this claim; in
particular, photospheric abundances from C, N, and O would be most useful.

\subsubsection{The CNO Cycle}
\label{sect:cnocycle}

Although most elements seem to follow an inverse FIP effect pattern,
another explanation is required for the high abundance of N (absolute or 
relative to Fe) in YY Men. Indeed, the N/Fe abundance ratio in Fig.~\ref{fig:ab_fip}
is much larger than the O/Fe abundance ratio, despite similar FIPs for O and N.
Some enhancement could be due to the inverse FIP effect, but we estimate this
effect to no more than $\sim 0.15$ dex, based on the [O/Fe] ratio.
Nitrogen enrichment in the photosphere of giant
stars due to dredge-up in the giant phase of material processed by the CNO cycle 
in the stellar interior is a candidate \citep[e.g.,][]{iben64,iben67}.
Significant mass transfer in close binaries can also reveal the composition
of the stellar interior. Such stars show X-ray spectra with an enhanced N/C abundance
ratio due to the CNO cycle \citep{schmitt02,drake03a,drake03b,schmitt04}.
Similar enhancement in V471 Tau was reported by \citet{drake03c} as evidence of
accreted material during the common enveloppe phase of the binary system. 
A high N/C ratio has been found in hot stars as well \citep[e.g.,][]{kahn01,mewe03}. 
YY Men, however, is a single giant, although it could have been formed in the merging of
a contact binary when one component moved into the giant phase \citep[e.g.,][]{rucinski90}

The abundance ratio is $\mathrm{[N/C] \sim +0.70^{+0.8}_{-0.4}}$ based on the \xmm\  fits
reported in Table~\ref{tab:meth}. \citet{schaller92} report an enhancement of $\mathrm{[N/C] =
+0.56}$ for a star of initial mass $M = 2.5~M_\sun$, and metallicity $Z = 0.02$ ($X =
0.68$, $Y=0.30$). Due to the uncertain origin of YY Men, such a number is
indicative only; however, it suggests that the high N coronal abundance
in fact reflects the photospheric composition of YY Men. Interestingly, the
prototype of its class, FK Com (G5 III), shows an unmixed \ion{N}{5}/\ion{C}{4}
ratio \citep{fekel93}. This is consistent with the line ratio of
\ion{N}{7}/\ion{C}{6} in X-rays \citep{gondoin02}. These ratios point to a
possible evolutionary difference between FK Com and YY Men, in which the former
is still in the Hertzsprung gap phase and has not undergone the He flash, and 
YY Men is a post-He flash clump giant with strong X-ray emission.

From \citet{schaller92}, the theoretical C/Ne ratio after the first dredge-up of
a giant is also depleted ($-0.16$~dex), whereas the O/Ne ratio essentially
remains constant ($-0.02$~dex). From our fits, $\mathrm{[C/Ne] \sim -0.5}$, and
$\mathrm{[O/Ne] \sim -0.5}$, lower than expected from evolutionary codes, 
but with large error bars. Nevertheless, we recall the enrichment  of coronal Ne
by a factor of about $+0.3$ to $+0.5$ dex  in active stars \citep[e.g.,][]{brinkman01,drake01} 
due to the inverse FIP effect. Such an enhancement lowers significantly the abundance
ratios relative to Ne, which can explain the lower  abundance ratios of [C/Ne] and
[O/Ne] than the evolutionary predictions.

The CNO cycle keeps the total number of C, N, and O atoms constant. In addition, for a
star of intermediate mass ($2.5~M_\sun$), the number of Ne atoms does not change
\citep{schaller92}. Thus, for a star with an initial solar 
composition, the ratio $\mathrm{(A_C+A_N+A_O)/A_{Ne}}$ should remain solar,
where $\mathrm{A_X}$ is the abundance number ratio of element X relative to H.
The solar photospheric values of \citet{grevesse98} are $\mathrm{A_C} = 3.31 \times 10^{-4}$,
$\mathrm{A_N} = 8.32 \times 10^{-5}$, $\mathrm{A_O} = 6.76 \times 10^{-4}$, and
$\mathrm{A_{Ne}} = 1.20 \times 10^{-4}$, yielding a ratio of $9.1$. 
Our best-fit abundances from \xmm\  data yield a ratio of $\sim 4.3$ with an estimated
uncertainty of $\pm 2.5$. Although the lower ratio again could cast some doubt on
the detection of CNO cycle in YY Men, we reiterate that coronal Ne is enhanced due to
the inverse FIP effect. If the photospheric Ne abundance in YY Men is two times lower
than in its corona, the above ratio would be in good agreement with the solar ratio.
Finally, although the above discussion assumes near-solar abundances in the stellar photosphere,
the observed Fe depletion in the photosphere \citep{randich93} argues
against a solar composition in YY Men in general.

\subsection{Emission Measure Distribution}
\label{sect:emd}

Figure~\ref{fig:emd_norm} displays the EMD of YY Men's corona derived from our
spectral fits to the \chan\  and \xmm\  data. We reiterate that we use $\mathrm{EMD}(T) / \Delta\log T$ for
our graphical representation of the EMD for methods 2 and 3,
whereas we plot the EM of each isothermal component for method 1. 
Although the plotted EMs for the latter method lie below the EMs per bin curves
for the two other methods, we emphasize that this is solely for plotting
purposes since the total EMs in YY Men are similar with the three method ($\log
\mathrm{EM_tot}$ (\cmcc) 
$\sim 55.3$ and $\sim 55.1$ during the Chandra and XMM-Newton observations, respectively).
The upper panel with a linear ordinate
emphasizes the dominant very hot plasma in YY Men. The lower panel, using a
logarithmic scale, reveals the weak EM at lower temperatures ($<20$~MK). Other FK Com-type stars show
similar EMDs with a dominant plasma at $T>20$~MK \citep{gondoin02,gondoin04}.

The various methods show consistently an EMD peaking around $20-40$~MK.
However, for the same data set, the derived EMD can be slightly different.
For example, the realization obtained from method 3 with the \chan\  spectra
show a smoother solution than the Chebychev solution (method 2). 
Method 3 starts with a smooth approximation and introduces structure
only as far as need to fit the line fluxes. A sufficiently good $\chi^2$ was reached before
any strong modification of the EMD at high-$T$ was introduced.
The good statistics and the excellent match between
the abundance ratios show that both EMDs are realistic and valid. 
We see no preference for either method, in particular in the light of 
the considerable systematic uncertainties and the consequent ill-posed spectral inversion problem.

We also note that the dominant plasma is at slightly lower temperatures ($20$~MK)
with \xmm\  than with \textit{Chandra} ($40$~MK). A larger EM for the highest-$T$
component is possible with \textit{XMM-Newton} but is unstable to
the fitting algorithm. It remains unclear whether YY Men was actually
cooler during the \textit{XMM-Newton} observation. Indeed, the statistical 
EM uncertainties with method 1 are smaller at high-$T$ than at mid-$T$ with 
\textit{Chandra} but  the trend is reversed with \textit{XMM-Newton} (Tab.~\ref{tab:meth}),
which suggests that, due to the spectral inversion problem, fits could have preferred a
somewhat cooler realization for the EMD of YY Men during the \xmm\  observation.

\subsubsection{Coronal Structures}
\label{sect:structures}

The lack of spatial resolution (in contrast to the Sun) does not allow
us to understand what structures are present in the corona of YY Men. We are 
restricted to loop formalisms \citep*[e.g.,][]{rosner78,mewe82,schrijver89,vandenoord97}.
The analytical formulae allow us to derive some 
information on the loop structures after placing certain assumptions 
(e.g. constant pressure, similar maximum temperature, constant loop
cross-section, etc). Discrepancies between the observed EMD and the
one predicted by loop models allow us to test some of the above
assumptions and modify them accordingly if necessary (e.g., by allowing 
loop expansion towards the top or the presence of distinct families 
of loops with different maximum temperatures). The high coronal
temperature (a few tens of MK) and the low surface gravity 
($\sim 0.01 g_{\odot}$) for YY Men imply that the pressure scale 
height $H_\mathrm{p}$ is large enough ($\sim 5 \times 10^{12}$~cm $= 6 R_\star$) 
to ensure constant pressure in most X-ray emitting loops.    

\citet*{argiroffi03} and \citet{scelsi04} have recently
published EMDs for the RS CVn binary Capella
observed by \chan\  and for the single rapidly rotating G0~III
giant 31 Com observed by \xmm, using the Markov-Chain Monte Carlo EMD
reconstruction method by \citet{kashyap98}. 
Figure~\ref{fig:emdcompare} compares the
EMD derived by us for YY Men with the EMDs of Capella and 31 Com
derived by these authors. In the figure we have used the EMD derived
from Chandra data using method 3 (and the graphical representation given in
Eq.~\ref{eq:emd}), since this method uses the same
bin width $\Delta\log T  = 0.1$ as used by \citet{argiroffi03} and
\citet{scelsi04}. 
Although the EMDs of Capella and 31 Com are similar, with 31 Com being 
only slightly hotter than Capella, the EMD  of YY Men peaks at higher $T$ 
and shows a shallow increase to the peak (Fig.~\ref{fig:emd_norm}). 

Figure~\ref{fig:emdcompareall} compares the EMD of YY Men
derived from \xmm\  data using method 1 with those derived with basically 
the same method  (RGS + MOS2 data and APEC code) by \citet{audard03} for a sample 
of RS CVn binaries of various activity levels (including Capella). 
We have added for comparison the results of a $3$-$T$ fit
to MOS2 data derived by \citet{scelsi04} for 31 Com
as well as the results of $2$-$T$ and $3$-$T$ fits to EPIC data (MOS+pn)
of two other FK Comae-type stars (V1794 Cyg and FK Com itself)
from \citet{gondoin02} and \citet{gondoin04}.
Figures~\ref{fig:emdcompare} and \ref{fig:emdcompareall} emphasize
the exceptional nature of YY Men (and of FK Com-type
giants in general) compared to other active stars in terms of high 
$T$ and high $EM$.

We can use the approach proposed by \citet{peres01} to infer
the properties of coronal loops from the observed EMD shape. If we
assume the loop model of \citet{rosner78}, we find that the shape of the emission 
measure distribution for a single loop does not depend on the length 
of the loop, but only on the loop maximum temperature $T_\mathrm{max}$, and 
its functional form is EMD($T$) $\propto T^{\beta}$ up to $T_\mathrm{max}$.
\citet{peres01} showed that, by grouping coronal loops by
their maximum temperature irrespective of length, it is possible
to describe the integrated properties of the corona, and the total
emission measure distribution, as the sum of the contributions 
of different families of loops, each characterized by a
different value of the maximum temperature $T_\mathrm{max}$. They showed, 
in particular, that under these assumptions the ascending
slope of the EMD($T$) of the whole corona is linked to $\beta$,
whereas the power-law index of the descending slope
is linked to the distribution of the maximum temperatures of different
classes of coronal loops. 

For YY Men we find a flatter EMD increase ($\beta \sim 1.5$ from method 3 and 
$\sim 2.15$ for method 2; see Fig.~\ref{fig:emd_norm}), 
closer to the solar slope. This could indicate that the heating is more uniform in 
the loops of YY Men than in Capella and 31 Com (which have $\beta \sim 5-6$;
\citealt{argiroffi03,scelsi04}) and that the loop
cross-section is approximately constant. We caution, however, that the method
used to derive the EMD for YY Men differ from those used by \citet{argiroffi03} and \citet{scelsi04}.
This could be responsible in part for the different slopes.
 A very steep gradient 
($T^{-6}$) is obtained at temperatures $> T_\mathrm{max}$ for Capella 
and YY Men and somewhat less (but with larger errors) for 31 Com. In the
loop formalism of \citet{peres01}, such steep slopes imply that
most of the loops have $T_\mathrm{max}$ equal to, or only slightly larger
than the peak value of the total emission measure distribution.

\subsubsection{Flare Heating}
\label{sect:heating}

The solar corona cools radiatively mostly around $1-2$~MK, with very weak
EM at higher $T$ \citep*[e.g.,][]{orlando00}. Stellar coronae display
higher coronal $T$, with FK Com-type stars at the upper end of the
active stars. A major problem of the solar-stellar connection studies is to
understand the mechanism at the origin of the high coronal $T$.
Coronal heating by flares is an attractive mechanism because
flares often display large temperatures \citep*[e.g.,][]{guedel99,franciosini01}. The
concept assumes a statistical contribution of flares to the energy budget
in solar and stellar coronae \citep[e.g.,][]{parker88}. Recent work on flaring
stars has shown that a continual superposition of flares (which follow a
power-law distribution in energy) could radiate sufficient energy to explain the
observed X-ray luminosity, suggesting that flares are significant contributors
to the coronal heating in the Sun and in stars
\citep*[e.g.,][]{guedel97,krucker98,audard99,audard00,parnell00,kashyap02,guedel03}.

The radiative cooling curve, $\Lambda(T)$, scaled to coronal
abundances in YY Men,
is flat in the $3-20$~MK range ($\Lambda \sim 1.2 \times 10^{-23}$~erg~s$^{-1}$~cm$^{3}$), 
in contrast to the curve for solar abundances 
(Fig.~\ref{fig:coolcurve}). The radiative loss time, $\tau_r = 3kT/(n_\mathrm{e} \Lambda(T))$,
is thus essentially directly proportional to the temperature (assuming constant
$n_\mathrm{e}$, see Sect.~\ref{sect:densities}) in this range. Figure~\ref{fig:radcooltime}
shows the radiative loss time as a function of $T$, using the cooling curve $\Lambda(T)$
in YY Men's corona, and assuming electron densities of $n_\mathrm{e} = 10^9$~\cmcc\  and
$n_\mathrm{e} = 10^{10}$~\cmcc. There is no evidence for high densities above $10^9-10^{10}$~\cmcc\  
in YY Men (see Sect.~\ref{sect:densities}). The radiative loss time is long ($> 0.1$~days), thus
we can expect that the occurrence time scale of stochastic flares is much shorter than the cooling time.

Consequently, the flare heating mechanism appears attractive in YY Men,
with frequent hot flares replenishing the corona on time scales shorter
than the radiative cooling time scale, thus keeping YY Men's corona hot by continual reheating. 
The lack of obvious flares in the \chan\  and \xmm\  light curves over
observing time spans of about a day, however, does not immediately support this hypothesis.
The \xmm\  EPIC MOS1 light curve (Fig.~\ref{fig:lc}) is on average $\mu \sim 1.3 \pm 0.1$~\cps,
corresponding to $L_\mathrm{X} \sim 1.6 \times 10^{32}$~\ergps. For a 3 $\sigma$ detection of
a flare, a minimum of $0.3$~\cps\  is required at peak, correspond to
$L^\mathrm{peak}_\mathrm{X} \geq 4 \times 10^{31}$~\ergps\  at YY Men's distance. 
A similar reasoning applies to the
\chan\  light curve. Such flares are at the
high end of the flare luminosity scale in stellar coronae and about 1,000 times stronger
than the strongest solar flares. Due to the 
large distance of YY Men, it is, therefore, no surprise that current detectors are unable to 
detect small-to-moderate flares in its corona since they are overwhelmed by fluctuations
due to Poisson statistics.

The above result suggests that if the flare heating mechanism is operating
in YY Men, the flare occurrence rate distribution in energy, $dN / dE \propto E^{-\alpha}$,
must be steep. \citet{guedel97} and \citet{guedel03} argued that the shape of
the EMD can provide information on the power-law index $\alpha$. In brief,
\citet{guedel03} obtained that the EMD is described by two power laws,
below and above a turnover, 

\begin{equation}
\text{EMD($T$)} \propto 
 \begin{cases}
   T^{2/\zeta},                                         & T \leq T_0,\\
   T^{-(b-\phi)(\alpha - 2\beta)/(1-\beta)+2b-\phi}, 	& T> T_0.
 \end{cases}
\end{equation}

 The turnover temperature, $T_0$, is related to the minimum peak luminosity, $L_\mathrm{min} = afT^{b-\phi}$,
from which the flare occurrence rate distribution needs to be integrated. The EMD rises
with an index $2/\zeta$. The index $\zeta$ corresponds to the index of the flare temperature
decay as a function of the plasma density, $T \propto n^\zeta$. Typical values are $\zeta = 0.5 - 2$
\citep*{reale93}. The other indices come from $\Lambda(T) = f T^{-\phi}$ (Fig.~\ref{fig:coolcurve}), 
the flare decay time $\tau \propto E^\beta$ ($E$ being the total radiative X-ray energy), 
and the flare peak emission measure, $\mathrm{EM} = a T^b$.
\citet{aschwanden99} reported $b = 7$ and $a = 0.6875$~\cmcc~K$^{-7}$  for the Sun (\citealt{feldman95,feldman96} quoted an exponential
function, $\mathrm{EM}(T) = 1.7 \times 10^{0.13T_6+46}$ \cmcc, which can be approximated with $b \sim 5-7$
in the range of interest; furthermore, \citealt{guedel04} obtains $b = 4.3 \pm 0.35$ using stellar flares).
We refer the reader to \citet{guedel03} for more details. 

In YY Men, $T_0$ occurs around $30$~MK, which corresponds
to $L_\mathrm{min} \sim 10^{28} - 10^{29}$~\ergps\  ($f = 2.28 \times 10^{-27}$~erg~s$^{-1}$~cm$^{3}$ 
and $\phi = - 1/2$, $b = 7$). Below $T_0$, $\mathrm{EMD}(T) \propto T^{2.15}$, implying $\zeta \sim 0.9$. 
Above $T_0$, $\mathrm{EMD}(T) \propto T^{-6}$.  For $\beta = 0$ (flare decay time $\tau$ 
independent of the flare energy), we obtain\footnote{Here, we use the continuous EMD with method 2.
Slightly lower indices $\alpha$ are found from the decreasing slope of the EMD with method 3; however, they
still are above the critical value of $2$.} $\alpha \sim 2.7$, weakly dependent of the exact choice of $b$ (lower $b$
values increase $\alpha$). For a weak energy dependence, $\beta = 0.25$, we get 
$\alpha = 2.6$, which still shows  that a wealth of flares with small energies 
could dominate the coronal heating in YY Men, but remain
undetected with the sensitivity of current detectors.
It is worthwhile to note that $L_\mathrm{min}$ is of the order of very large solar flares, implying
that, while smaller flares could be present in YY Men, they are not needed to explain
the X-ray emission of YY Men.

\subsection{Line Broadening}
\label{sect:doppler}

Excess line width is observed with \textit{Chandra} in \ion{Ne}{10} $\lambda
12.13$~\AA, and, albeit at lower significance, in \ion{Si}{14} $\lambda 6.18$~\AA, \ion{O}{8} $\lambda 18.97$~\AA, 
\ion{Mg}{12} $\lambda 8.42$~\AA, and possibly \ion{N}{7} $\lambda 24.78$~\AA.
We give in Table~\ref{tab:excesswidth} the line fluxes and the width measurements in excess of the
instrumental width together
with their 68\% confidence ranges.
Other emission lines, however, do not show evidence of widths in excess of the instrumental profile.
We propose that the observed line broadening has a Doppler thermal broadening origin. In a
Maxwellian velocity distribution of particles with a temperature $T$, thermal movements 
broaden the natural frequency (or wavelength) width of a transition centered at $\nu_0$,
and produce a line profile function $\phi(\nu)$ 
\citep[e.g.,][]{rybicki79},
\begin{equation}
\phi(\nu) = \frac{1}{\Delta\nu_D \sqrt{\pi}} e^{-(\nu - \nu_0)^2 / (\Delta\nu_D)^2},
\end{equation}
where the Doppler thermal width is obtained from
\begin{equation}
\frac{\Delta\nu_D}{\nu_0} = \frac{\Delta\lambda_D}{\lambda_0} = \sqrt{\frac{2kT}{Mc^2}},
\end{equation}
with $M$ as the atomic mass of the element. The observed broadening, $\sigma$, is related
to the Doppler thermal width,
\begin{equation}
\sigma = \frac{\Delta\lambda_D}{\sqrt{2}} = \frac{\lambda_0}{c} \sqrt{\frac{kT}{M}}.
\label{eq:width}
\end{equation}

For an isothermal plasma of $40$~MK, the Doppler broadening is 18.7~m\AA\  for 
\ion{C}{6}, 12.7~m\AA\   for \ion{N}{7}, 9.1~m\AA\   for \ion{O}{8}, 5.2~m\AA\  for 
\ion{Ne}{10}, 3.2~m\AA\  for \ion{Mg}{12}, 2.2~m\AA\  for \ion{Si}{14}, 1.6~m\AA\   for \ion{S}{16}, 
1.1~m\AA\  for \ion{Ar}{18}, and 0.9~m\AA\  for \ion{Ca}{20} Ly$\alpha$ lines. For Fe lines,
the broadening is 2.6~m\AA\  and 3.8~m\AA\  at 10~\AA\  and 15~\AA, respectively.

Although the above isothermal approximation works for the strongly peaked EMD
of YY Men, a more thorough analysis can be given. Based on line emissivities
and the derived EMD, we determine from which temperature the photon luminosity 
arises. The photon luminosity of line $i$, $\mathcal{L}_i$, is determined as
\begin{equation}
\mathcal{L}_i= \int \Lambda_i (T) \varphi(T) dT
\end{equation}
where $\Lambda_i (T)$ is the photon emissivity ($\mathrm{ph~cm^3~s^{-1}}$) of line $i$ from APEC, and $\varphi(T)$
is as in Eq.~\ref{eq:dem}. Therefore, a graphical representation of the photon luminosity
distribution is $L_i(T) = \Lambda_i(T) \varphi(T) T \Delta\log T  = \Lambda_i(T) \times \mathrm{EMD}(T)$,
which is shown in Figure~\ref{fig:lumline} (top panel) for \ion{O}{8} and \ion{Ne}{10} Ly$\alpha$ lines, using
abundances and EMD from the \chan\   method 2 fit. \emph{Despite the line maximum formation temperatures being at
$3$~MK and $6$~MK, respectively,  most photons come from $T$ bins in the range of $6-15$~MK and $20-60$~MK!}
Also, line photons in bin $T$ are broadened by a specific square width, $\sigma^2_i(T)$ (Eq.~\ref{eq:width}).
The distribution of square widths weighted by the photon luminosity,
$<\sigma^2_i(T)> = \sigma^2_i(T) \times L_i(T) / \mathcal{L}_i$,
shows the contribution of each $T$ bin to the square width of the line (Fig.~\ref{fig:lumline}, bottom panel).
The distribution shows that the width is dominated by the $30-50$~MK plasma component.

Since we used the \chan\  calibration with the latest LSF profile, we believe that
the measured excess widths are not related to inaccuracies in the instrumental profile.
We tested further the robustness of our results, e.g., in \ion{Ne}{10} and \ion{Si}{14}.
We simultaneously fitted the HEG negative and positive first-order line profiles
with Gaussian functions, with the wavelengths left free to vary in both spectra (to allow for 
wavelength calibration inaccuracies between the spectra), and letting the width and flux free as well, 
however linking the two parameters in both sides.
Then, an instrumental profile was fitted to the emission line. The latter wavelengths and 
flux were used to simulate 10,000 realizations of a similar emission line. Each realization
was then fitted with a Gaussian function with free width. This procedure aims 
to obtain the 
fraction of best-fit Gaussian widths larger or equal to the detected excess width, which
provides us with an estimate of the detection significance. 

The \ion{Ne}{10} Ly$\alpha$ line showed a width of $\sigma = 6.2$~m\AA\  with a 90\% confidence range 
of $[4.4,7.8]$~m\AA. The moments of the widths of the 10,000 Monte-Carlo simulations show an average of
1.2~m\AA, with a standard deviation of 1.0~m\AA. More interestingly, the maximum width
fitted from a model with an instrumental profile is 5.3~m\AA, \emph{lower} than the best-fit
excess width. Essentially, this shows that there is a probability less than $10^{-4}$ that the
detected excess width in the \ion{Ne}{10} Ly$\alpha$ line is spurious. Taking into account the
90\% confidence range, there are only 66 occurrences out of 10,000 where a best-fit width larger than or
equal to the lower limit of the range, i.e., 4.4~m\AA\  was spuriously measured, although we
emphasize that the corresponding line fluxes are weak.
Figure~\ref{fig:NeXLya} (left) shows the $\Delta C = 1$ and $\Delta C = 2.71$
confidence maps in the line flux/sigma space together with the best-fit values of the 10,000
Monte-Carlo simulations. Figure~\ref{fig:NeXLya} (right) shows the observed line
profile, the instrumental profile, and the best-fit Gaussian model.

Similarly, the \ion{Si}{14} Ly$\alpha$ line showed some evidence of an excess width. However,
the poor signal-to-noise ratio and the smaller spectral power at 6~\AA\  (essentially half the 
power at 12~\AA) diminishes the significance of the detection. We applied the same procedure
as for the \ion{Ne}{10} line; however, we rebinned the spectra by a factor of two (bin size of
5~m\AA) to increase the number of counts per bin. The best-fit width was 4.4~m\AA\  with a
90\% confidence range  of $[2.4,6.2]$~m\AA. The significance
is lower (209 out of 10,000 values are larger than or equal to the best-fit width, and 2,107 out of
10,000 for the lower limit); however, the loci of the Gaussian widths of the Monte-Carlo 
realizations suggests that the detection may be real (Fig.~\ref{fig:SiXIVLya},
left panel). Figure~\ref{fig:SiXIVLya} (right) again shows the observed line
profile, the instrumental profile, and the best-fit Gaussian model.

It appears that a broadening smaller than $\sim 3$~m\AA\  proves difficult to measure,
which could explain the absence of broadening in the short wavelengths, e.g., in the
\ion{S}{15}, \ion{Ar}{18}, \ion{Ca}{20} lines. On the other hand, it cannot
explain the non-detections in, e.g., the Fe lines around $15$~\AA\  and the
upper limit of \ion{N}{7} at $24.7$~\AA. We believe that the poor signal-to-noise
ratio of these lines against the underlying continuum is at the origin of this
discrepancy. Few counts are measured, in contrast to the strong signal of the
\ion{Ne}{10} Ly$\alpha$ line. We note that, despite the strong flux of the
\ion{O}{8} Ly$\alpha$ line, interstellar absorption and effective areas both
conspired to reduce the amount of detected counts significantly.

Although the above interpretation focused on Doppler thermal broadening only, we note
that line broadening could be due to the stellar rotation as well, in part or completely.
Indeed, \citet{ayres98} found evidence in UV of broad line profiles of the fastest rotating
gap giants (YY Men was not part of the sample, but it included FK Com). The broadening 
suggested the presence of emission sources in the transition zone at heights of $\sim R_\star$ 
above the photosphere. In addition, \citet{chung04} found excess line broadening in the \chan\  X-ray spectrum
of Algol which they interpreted as rotational broadening from a radially extended coronae at
temperatures below 10 MK and with a scale height of order the stellar radius.

Measurements of the projected equatorial rotational velocity in YY Men indicates that 
$v \sin i = 45$~\kms\  \citep{piskunov90}, which implies a maximum wavelength shift of
\begin{equation}
\Delta\lambda_\mathrm{max}  = \frac{\lambda}{c} v \sin i = 1.82 \times \Big(\frac{\lambda}{12.134}\Big) \textrm{~m\AA}
\end{equation}
with $\lambda$ in \AA. 
Therefore, for structures at the stellar surface, rotational broadening is smaller than 
Doppler broadening in the \textit{Chandra} HETGS and \textit{XMM-Newton} RGS wavelength range, and therefore 
it probably does not explain the observed line broadening. However, if coronal X-ray emitting
material is high above the surface (e.g., at the pressure scale height, $H_\mathrm{p} = 6 R_\star$), 
it could produce significant broadening visible in the X-rays as well. If interpreted
as purely rotational broadening, the excess width for \ion{Ne}{10}, $\sigma \sim 6$~m\AA\  (Tab.~\ref{tab:excesswidth}),
implies a velocity of about $200$~\kms, using the inclination angle of $65\arcdeg$ \citep{piskunov90}.
The latter authors also found no polar spots in YY Men but equatorial belts. Consequently, the excess width suggests
that the coronal X-ray emitting material lies at the equator at heights of about $3 R_\star$ above the
surface, i.e., below the pressure scale height. We note that at the wavelength of \ion{Si}{14},
a larger velocity ($300$~\kms) is required to match the observed excess width, i.e., a higher 
altitude ($6 R_\star$). Although rotational broadening remains a possible explanation, we favor
Doppler thermal broadening instead because of the dominant very hot plasma temperature in YY Men
(see Fig.~\ref{fig:lumline}, bottom panel), in comparison, e.g., to the lower plasma temperature in Algol 
where rotational broadening was the preferred scenario of \citet{chung04}.


\subsection{Densities}
\label{sect:densities}

The \chan\  and \xmm\  grating spectra cover ranges that include transitions whose intensities
are density-sensitive. In particular, line ratios of the forbidden ($f$) to intercombination
($i$) lines of He-like transitions are most useful since they are sensitive to plasma
electron densities covering those found in stellar coronae ($10^9-10^{11}$~\cmcc; e.g.,
\citealt*{ness02,ness03a,ness04,testa04a}).

The very hot corona of YY Men is, however, problematic to derive electron densities from
He-like transitions since the latter are faint and of low contrast against the strong underlying 
continuum (e.g., Fig.~\ref{fig:chandra_bfit_meth1a}--\ref{fig:chandra_bfit_meth1c}). Indeed, most He-like triplets where the 
\chan\  HETGS
and \xmm\  RGS effective areas and spectral powers are large enough (\ion{Si}{13}, \ion{Mg}{11},
\ion{Ne}{9}, \ion{O}{7}) are formed at $T \leq 10$~MK, where the EM is much lower ($\sim 1 $~dex)
than the peak EM (Fig.~\ref{fig:emd_norm}). In addition, blending is frequent (e.g., contamination
of \ion{Ne}{9} by Fe lines; \citealt{ness03b}).  

We extracted individual
line fluxes for \ion{Si}{13}, \ion{Mg}{11}, \ion{Ne}{9}, and \ion{O}{7} from the \chan\  
spectrum (Tab.~\ref{tab:chanfluxes}) because HETGS offers the best spectral resolution.
We used the \textit{Chandra} HEG1 and MEG1 data simultaneously to fit  emission lines individually
together with a ``line-free'' continuum (see Sect.~\ref{sect:method3}). 
No line fluxes of He-like triplets were obtained
from the \xmm\  RGS spectrum because of the lower spectral resolution for \ion{Si}{13} and
\ion{Mg}{11}, of the Fe blending in \ion{Ne}{9}, and of the low signal-to-noise
ratio of the \ion{O}{7} and \ion{N}{6} triplets (Fig.\ref{fig:xmmspec}). 

Unfortunately, the lack of signal of the He-like triplets in YY Men produced large
confidence ranges\footnote{We derived confidence ranges from a grid of line fluxes,
allowing the statistics to reach $\Delta C = 1$. Our method accounts for uncertainties
in the determination of the continuum, for $N_\mathrm{H}$, and for calibration. 
We avoided calculating uncertainties from the square-root of the number of counts 
in the emission line since this method, though producing smaller
uncertainties, does not take into account the effects mentioned above.}. The
derived $R$ ratios ($R = f / i$) have uncertainties that do not allow us to
constrain significantly the electron densities. However, the $R$ ratios based on
the best-fit fluxes give no indication of high ($>10^{10}$~\cmcc) densities in YY Men.

\subsection{Opacities}
\label{sect:opacities}

Our models assumed an optically thin plasma in YY Men's corona. A systematic
study of stellar coronae with \xmm\  and \chan\  by \citet{ness03c} showed that
stellar coronae are generally optically thin over a wide range of activity
levels and average coronal temperatures. However, a recent study of the
Ly$\alpha$/Ly$\beta$ line ratio in active RS CVn binaries by \citet{testa04b} 
found evidence of opacity effects, suggesting that opacity measurements by means
of line ratios of \ion{Fe}{17} could be hampered in active stellar coronae 
which are strongly Fe-depleted \citep[e.g.,][]{audard03}.

We obtained estimates of the optical depths, $\tau$, from line ratios, using
the  ``escape-factor'' model designed by \citet{kaastra95} for $\tau \leq 50$,
\begin{equation}
\frac{R}{R_\mathrm{0}} = \frac{1}{1 + 0.43\tau},
\end{equation}
where $R$ is the ratio of the flux of an opacity-sensitive
line to the flux of an opacity-insensitive line, e.g., the
ratio of \ion{Fe}{17} $\lambda 15.01$~\AA\  to \ion{Fe}{17} $\lambda 15.26$~\AA\
(Tab.~\ref{tab:chanfluxes}), and $R_0$ is the ratio for an optically thin
plasma. We note that the optically thin ratio of the above lines is subject to 
debate. Whereas theoretical codes range from $3.0-4.7$ \citep{bhatia92}, laboratory
measurements obtain lower ratios ($2.8-3.2$; \citealt{brown98,laming00}). Despite
these uncertainties, $F(15.01)/F(15.26) \sim 2.7$ is compatible with no significant
optical depth in YY Men's corona (although uncertainties formally suggest a possible
optical depth, however with $\tau \leq 3$). The ratio of the \ion{Fe}{17} $\lambda 15.01$~\AA\  
to \ion{Fe}{17} $\lambda 16.78$~\AA, a measure of optical depth as well, is
close to the expected values for an optically thin plasma \citep{smith01,doron02,ness03c}.

In view of the results by \citet{testa04b}, opacity effects could possibly be
better measured in the Ly$\alpha$/Ly$\beta$ ratios of abundant elements. As
derived from Table~\ref{tab:chanfluxes}, the line ratio for \ion{Ne}{10} is
consistent with the theoretical ratio in APEC. The ratio for \ion{O}{8} is 
slightly lower, but we argue that contamination by \ion{Fe}{18} could reduce the
ratio artificially. Ly$\alpha$/Ly$\beta$ line ratios
of, e.g., \ion{N}{7}, \ion{Mg}{12}, \ion{Si}{14} look consistent with the
theoretical ratios as well, although the weakness of the Ly$\beta$ lines makes
the accurate measurement of such ratios difficult. 
In conclusion, there is no strong support for a significant optical depth in 
the corona of YY Men, in line with the study by \citet{ness03c}.

\section{Summary and Conclusions}
\label{sect:conclusions}

In this paper, we have presented our analysis of the X-ray emission of the FK Com-type giant star, 
YY Men, observed by \chan\  HETGS and \xmm. Highly ionized Fe lines, H-like transitions, and a strong 
underlying continuum in the high-resolution X-ray spectra
reveal a dominant very hot plasma. We used three different methods to derive the
EMD and coronal abundances and all three show a strong peak EM around $20-40$~MK,
about a dex above EMs at lower $T$ (Fig.~\ref{fig:emd_norm}). We compared the EMD
with other giants and active stars (Figs.~\ref{fig:emdcompare} and \ref{fig:emdcompareall})
to emphasize the exceptional coronal behavior of YY Men given its high $T$ and EM.
Such a hot plasma produces thermal broadening at the level detectable by \chan. Indeed, we
measured line broadening in several lines, which we interpreted as predominantly Doppler thermal broadening
(Sect.~\ref{sect:doppler}). 

YY Men was about two times brighter, and possibly slightly hotter, during the 
\chan\  observation than during the \xmm\  observation. 
No evidence for flares, or small-scale variations down to $\sim
300$~s was found in the \xmm\  and \chan\  light curves (Sect.~\ref{sect:lc}). Nevertheless, the
absence of variability does not imply absence of flares as the latter
need peak luminosities $L^\mathrm{peak}_\mathrm{X} \geq 10^{31}$~\ergps\  
to be detected with current detectors. We interpreted the shape of the EMD (Fig.~\ref{fig:emd_norm})
of YY Men's corona with two different formalisms. The first one infers the properties
of coronal loops from the EMD shape (Sect.~\ref{sect:structures}). From the
steep slope of the EMD at high $T$, we derived
that most of the loops in YY Men's corona have their maximum $T$ equal to or slightly
above $30$~MK. The second formalism makes use of the EMD in the context of coronal heating 
(Sect.~\ref{sect:heating}). We argued that a statistical ensemble of flares 
distributed in energy with a steep power law, $dN / dE \propto E^{-2.7}$, down to $L_\mathrm{min} 
\sim 10^{28} - 10^{29}$~\ergps\  could explain the decrease of the EMD at high
$T$ and the X-ray emission of YY Men. 
The steep index of the power law suggests that small flares could contribute most to the 
coronal heating in YY Men.

There is a marked depletion of low-FIP elements with respect to high-FIP
elements in YY Men's corona, suggesting an inverse FIP effect like in most
active RS CVn binaries (Sect.~\ref{sect:fip}). The lack of determinations of
photospheric abundances for individual elements except Fe  does not
allow us to determine whether the FIP-related abundance bias still holds when
coronal abundances are compared to the stellar photospheric composition instead 
of the solar. However, a photospheric [Fe/H] abundance found in the literature
indicates that the coronal Fe abundance is actually depleted. The high N
abundance found in YY Men's corona is interpreted as a signature of the CNO
cycle due to dredge-up in the giant phase (Sect.~\ref{sect:cnocycle}).

The low-signal-to-noise ratios in the He-like triplets prevented us
from obtaining definitive values for the electron densities (Sect.~\ref{sect:densities}). 
In addition, no significant optical depth was measured from line ratios
(Sect.~\ref{sect:opacities}). 
 
In conclusion, FK Com-type giants emit strong X-rays and contain the hottest coronal 
plasmas found in the large population of stars with magnetic activity.
Their study is important to understand the connection between the Sun and stars,
as they provide the most extreme conditions (e.g., large radius, rapid rotation, 
high coronal temperature) in magnetically active stars.

\acknowledgments

We acknowledge support from SAO grant GO2-3016X, from NASA grant NAG5-13553, 
from the Swiss National Science
Foundation (grants 20-58827.99 and 20-66875.01), from the UK Particle Physics and
Astronomy Research Council (PPARC), and from NASA to Columbia University for 
\textit{XMM-Newton} mission support
and data analysis. Based in part on observations obtained with \textit{XMM-Newton}, 
an ESA science mission with instruments and contributions directly funded by
ESA Member States and NASA. We thank an anonymous referee for useful comments
and suggestions that improved the content of this paper. M.~A. warmly thanks John Houck and
David Huenemoerder for their help and support with ISIS and the \textit{Chandra}
data, Jean Ballet for useful information on EPIC, Randall Smith for
discussions about the APEC database.  This paper profited from Stephen Drake, Rolf Mewe, and Anton
Raassen for useful comments about the manuscript. We thank Costanza Argiroffi and Luigi
Scelsi for providing figures for our comparison of YY Men with other stars.
Finally, we wish to dedicate this paper to the late Rolf Mewe who passed away during
the completion of this work, and whose presence and friendship will be greatly missed.


\begin{deluxetable}{lllll}
\tabletypesize{\footnotesize}
\tablecolumns{5}
\tablewidth{0pc}
\tablecaption{Observation Log for \chan\  and \xmm\label{tab:log}}
\tablehead{
\multicolumn{2}{c}{Instrument (Mode / Filter)} & \colhead{Start} &
\colhead{Stop} & \colhead{Exposure (ks)\tablenotemark{a}}}
\startdata
\multicolumn{5}{c}{\chan\  (ObsId 200165)}\\
 \noalign{\vskip .8ex}%
\multicolumn{2}{l}{ACIS-S/HETG} & 2002 Feb 1 23:57:01 UT & 2002 Feb 2
21:13:46 UT & 74.2\\
 \noalign{\vskip .8ex}%
\hline
 \noalign{\vskip 1.5ex}%
\multicolumn{5}{c}{\xmm\  (Rev 334, ObsId 0137160201)}\\
 \noalign{\vskip .8ex}%
\multicolumn{2}{l}{RGS1} &  2001 Oct 5 12:16:31 UT & 2001 Oct 6
12:50:24 UT & 86.3\\
\multicolumn{2}{l}{RGS2} &   2001 Oct 5 12:16:31 UT & 2001 Oct 6
12:52:20 UT & $84.1$\\
MOS1 &(Small Window / Thick) & 2001 Oct 5 12:23:05 UT & 2001 Oct 6
12:43:09 UT & $80.8$\\
MOS2 &(Small Window / Thick) & 2001 Oct 5 12:23:05 UT & 2001 Oct 6
12:43:56 UT & $80.8$\tablenotemark{b}\\
pn\tablenotemark{c}  & \nodata & \nodata & \nodata \\
OM & (Fast / $U$) & 2001 Oct 5 12:20:24 UT & 2001 Oct 6
12:04:34 UT & $84.9$\tablenotemark{d}
\enddata
\tablenotetext{a}{Usable exposure after filtering.}
\tablenotetext{b}{MOS1 data only were used.}
\tablenotetext{c}{pn camera off-line.}
\tablenotetext{d}{Fast window off-center; fixed pattern noise.}
\end{deluxetable}

\begin{deluxetable}{lllllll}
\tabletypesize{\footnotesize}
\tablecolumns{7}
\tablewidth{0pc}
\tablecaption{Spectroscopic Results from Various Methods with \textit{Chandra} and \textit{XMM-Newton}\label{tab:meth}}
\tablehead{
\colhead{} & \multicolumn{3}{c}{\textit{Chandra}} & \colhead{} & \multicolumn{2}{c}{\textit{XMM-Newton}}\\
\cline{2-4} \cline{6-7} \\ 
\colhead{\makebox[3.0cm][l]{Parameter}}  &  \colhead{M1\tablenotemark{a}} &   \colhead{M2\tablenotemark{a}} &  \colhead{M3} &  \colhead{} &
	 \colhead{M1\tablenotemark{a}} &  \colhead{M2\tablenotemark{a}}}
\startdata
$\log \mathrm{N_H}$ (\cmsq)\dotfill    & $20.87^{+0.05}_{-0.06}$ 			& $20.88^{+0.04}_{-0.04}$ 			& $20.89^{+0.02}_{-0.02}$			& {} & $20.83^{+0.03}_{-0.03}$		& $20.82^{+0.03}_{-0.03}$	   \\
$\mathrm{[C/Fe]}$\dotfill               & \nodata				     	& \nodata				        & \nodata					& {} & $+0.12^{+0.33}_{-0.72}$		& $+0.01^{+0.31}_{-0.74}$	   \\
$\mathrm{[N/Fe]}$\dotfill   	       & $+0.99^{+0.24}_{-0.30}$ 			& $+0.92^{+0.24}_{-0.32}$   			& $+0.79^{+0.17}_{-0.28}$			& {} & $+0.77^{+0.18}_{-0.18}$		& $+0.77^{+0.15}_{-0.15}$	   \\
$\mathrm{[O/Fe]}$\dotfill   	       & $+0.15^{+0.14}_{-0.15}$ 			& $+0.12^{+0.12}_{-0.12}$			& $+0.08^{+0.09}_{-0.12}$			& {} & $+0.08^{+0.11}_{-0.10}$		& $+0.11^{+0.08}_{-0.08}$	   \\
$\mathrm{[Ne/Fe]}$\dotfill  	       & $+0.52^{+0.10}_{-0.10}$ 			& $+0.52^{+0.08}_{-0.07}$			& $+0.52^{+0.14}_{-0.21}$ 			& {} & $+0.55^{+0.13}_{-0.11}$		& $+0.51^{+0.10}_{-0.09}$	   \\
$\mathrm{[Mg/Fe]}$\dotfill  	       & $-0.03^{+0.12}_{-0.13}$ 			& $-0.03^{+0.11}_{-0.10}$			& $-0.01^{+0.07}_{-0.09}$ 			& {} & $+0.12^{+0.17}_{-0.17}$		& $+0.09^{+0.15}_{-0.16}$	   \\
$\mathrm{[Al/Fe]}$\dotfill  	       & $+0.06^{+0.37}_{-2.09}$ 			& $+0.08^{+0.35}_{-1.27}$			& \nodata		 			& {} & \nodata				& \nodata	     	   	   \\
$\mathrm{[Si/Fe]}$\dotfill  	       & $+0.00^{+0.11}_{-0.12}$ 			& $-0.01^{+0.11}_{-0.10}$			& $-0.01^{+0.10}_{-0.13}$ 			& {} & $-0.11^{+0.18}_{-0.19}$		& $-0.13^{+0.14}_{-0.22}$	   \\
$\mathrm{[S/Fe]}$\dotfill   	       & $-0.10^{+0.17}_{-0.20}$ 			& $-0.08^{+0.16}_{-0.18}$			& $-0.09^{+0.11}_{-0.15}$ 			& {} & $-0.45^{+0.29}_{-0.51}$		& $-0.54^{+0.30}_{-0.66}$	   \\
$\mathrm{[Ar/Fe]}$\dotfill  	       & $+0.42^{+0.21}_{-0.29}$ 			& $+0.46^{+0.21}_{-0.25}$			& $< + 0.74$		 			& {} & $<+0.14$				& $<+0.04$	     	   	   \\
$\mathrm{[Ca/Fe]}$\dotfill  	       & $+0.00^{+0.36}_{-1.49}$ 			& $+0.04^{+0.35}_{-1.33}$			& $+0.40^{+0.22}_{-0.49}$ 			& {} & $<+0.35$				& $<+0.28$			   \\
$\mathrm{[Fe/H]}$\dotfill  	       & $-0.52^{+0.05}_{-0.05}$ 			& $-0.57^{+0.03}_{-0.04}$			& $-0.53^{+0.06}_{-0.07}$ 			& {} & $-0.68^{+0.05}_{-0.07}$		& $-0.64^{+0.04}_{-0.05}$	   \\
$\log T_1$ (K)\dotfill  	       & $6.86^{+0.03}_{-0.05}$ 			& $:6.075$\tablenotemark{b}		        & $:6.0$\tablenotemark{b}			      & {} & $6.45^{+0.06}_{-0.08}$	      & $:6.075$\tablenotemark{b}    \\
$\log T_2$ (K)\dotfill  	       & $7.16^{+0.04}_{-0.02}$ 			& $:6.225$\tablenotemark{b}		        & $:6.1$\tablenotemark{b}			      & {} & $6.91^{+0.03}_{-0.02}$	      & $:6.225$\tablenotemark{b}    \\
$\log T_3$ (K)\dotfill  	       & $7.60^{+0.02}_{-0.03}$ 			& $:6.375$\tablenotemark{b}		        & $:6.2$\tablenotemark{b}			      & {} & $7.26^{+0.04}_{-0.04}$	      & $:6.375$\tablenotemark{b}    \\
$\log T_4$ (K)\dotfill  	       & \nodata					& $:6.525$\tablenotemark{b}		        & $:6.3$\tablenotemark{b}			      & {} & $7.66^{+0.16}_{-0.06}$	      & $:6.525$\tablenotemark{b}    \\
$\log T_5$ (K)\dotfill  	       & \nodata					& $:6.675$\tablenotemark{b}		        & $:6.4$\tablenotemark{b}			      & {} & \nodata			      & $:6.675$\tablenotemark{b}    \\
$\log T_6$ (K)\dotfill  	       & \nodata					& $:6.825$\tablenotemark{b}		        & $:6.5$\tablenotemark{b}			      & {} & \nodata			      & $:6.825$\tablenotemark{b}    \\
$\log T_7$ (K)\dotfill  	       & \nodata					& $:6.975$\tablenotemark{b}		        & $:6.6$\tablenotemark{b}			      & {} & \nodata			      & $:6.975$\tablenotemark{b}    \\
$\log T_8$ (K)\dotfill  	       & \nodata					& $:7.125$\tablenotemark{b}		        & $:6.7$\tablenotemark{b}			      & {} & \nodata			      & $:7.125$\tablenotemark{b}    \\
$\log T_9$ (K)\dotfill  	       & \nodata					& $:7.275$\tablenotemark{b}		        & $:6.8$\tablenotemark{b}			      & {} & \nodata			      & $:7.275$\tablenotemark{b}    \\
$\log T_{10}$ (K)\dotfill	       & \nodata					& $:7.425$\tablenotemark{b}		        & $:6.9$\tablenotemark{b}			      & {} & \nodata			      & $:7.425$\tablenotemark{b}    \\
$\log T_{11}$ (K)\dotfill	       & \nodata					& $:7.575$\tablenotemark{b}		        & $:7.0$\tablenotemark{b}			      & {} & \nodata			      & $:7.575$\tablenotemark{b}    \\
$\log T_{12}$ (K)\dotfill	       & \nodata					& $:7.725$\tablenotemark{b}		        & $:7.1$\tablenotemark{b}			      & {} & \nodata			      & $:7.725$\tablenotemark{b}    \\
$\log T_{13}$ (K)\dotfill	       & \nodata					& $:7.875$\tablenotemark{b}		        & $:7.2$\tablenotemark{b}			      & {} & \nodata			      & $:7.875$\tablenotemark{b}    \\
$\log T_{14}$ (K)\dotfill	       & \nodata					& \nodata					& $:7.3$\tablenotemark{b}			      & {} & \nodata			      & \nodata 			 \\
$\log T_{15}$ (K)\dotfill	       & \nodata					& \nodata					& $:7.4$\tablenotemark{b}			      & {} & \nodata			      & \nodata 			 \\
$\log T_{16}$ (K)\dotfill	       & \nodata					& \nodata					& $:7.5$\tablenotemark{b}			      & {} & \nodata			      & \nodata 			 \\
$\log T_{17}$ (K)\dotfill	       & \nodata					& \nodata					& $:7.6$\tablenotemark{b}			      & {} & \nodata			      & \nodata 			 \\
$\log T_{18}$ (K)\dotfill	       & \nodata					& \nodata					& $:7.7$\tablenotemark{b}			      & {} & \nodata			      & \nodata 			 \\
$\log T_{19}$ (K)\dotfill	       & \nodata					& \nodata					& $:7.8$\tablenotemark{b}			      & {} & \nodata			      & \nodata 			 \\
$\log T_{20}$ (K)\dotfill	       & \nodata					& \nodata					& $:7.9$\tablenotemark{b}			      & {} & \nodata			      & \nodata 			 \\
$\log T_{21}$ (K)\dotfill	       & \nodata					& \nodata					& $:8.0$\tablenotemark{b}			      & {} & \nodata			      & \nodata 			 \\
$\log\mathrm{EM}_1$ (\cmcc)\dotfill    & $53.95^{+0.12}_{-0.15}$ 			& $51.06 \pm 0.17$\tablenotemark{c}		& $52.07 \quad (51.67 - 51.87)\tablenotemark{d}$      	& {} & $53.61^{+0.17}_{-0.29}$  	& $52.33 \pm 0.34$\tablenotemark{c}\\
$\log\mathrm{EM}_2$ (\cmcc)\dotfill    & $54.27^{+0.14}_{-0.21}$ 			& $52.86 \pm 0.12$\tablenotemark{c}		& $52.27 \quad (52.07 - 52.27)\tablenotemark{d}$      	& {} & $54.16^{+0.09}_{-0.07}$  	& $53.24 \pm 0.25$\tablenotemark{c}\\
$\log\mathrm{EM}_3$ (\cmcc)\dotfill    & $55.22^{+0.02}_{-0.02}$ 			& $52.01 \pm 0.12$\tablenotemark{c}		& $52.47 \quad (52.45 - 52.62)\tablenotemark{d}$      	& {} & $54.86^{+0.04}_{-0.04}$  	& $52.52 \pm 0.21$\tablenotemark{c}\\
$\log\mathrm{EM}_4$ (\cmcc)\dotfill    & \nodata        			     	& $52.04 \pm 0.10$\tablenotemark{c}		& $52.67 \quad (52.59 - 52.73)\tablenotemark{d}$      	& {} & $54.68^{+0.14}_{-0.15}$  	& $52.56 \pm 0.21$\tablenotemark{c}\\
$\log\mathrm{EM}_5$ (\cmcc)\dotfill    & \nodata        			     	& $52.89 \pm 0.12$\tablenotemark{c}		& $52.83 \quad (52.91 - 53.04)\tablenotemark{d}$      	& {} & \nodata  			& $53.16 \pm 0.21$\tablenotemark{c}\\
$\log\mathrm{EM}_6$ (\cmcc)\dotfill    & \nodata        			     	& $53.67 \pm 0.13$\tablenotemark{c}		& $52.91 \quad (53.54 - 53.62)\tablenotemark{d}$      	& {} & \nodata  			& $53.67 \pm 0.24$\tablenotemark{c}\\
$\log\mathrm{EM}_7$ (\cmcc)\dotfill    & \nodata        			     	& $53.91 \pm 0.07$\tablenotemark{c}		& $52.94 \quad (53.48 - 53.97)\tablenotemark{d}$      	& {} & \nodata  			& $53.92 \pm 0.23$\tablenotemark{c}\\
$\log\mathrm{EM}_8$ (\cmcc)\dotfill    & \nodata        			     	& $53.82 \pm 0.13$\tablenotemark{c}		& $53.06 \quad (54.25 - 54.34)\tablenotemark{d}$      	& {} & \nodata  			& $54.13 \pm 0.24$\tablenotemark{c}\\
$\log\mathrm{EM}_9$ (\cmcc)\dotfill    & \nodata        			     	& $53.93 \pm 0.12$\tablenotemark{c}		& $53.23 \quad (54.33 - 54.23)\tablenotemark{d}$      	& {} & \nodata  			& $54.47 \pm 0.22$\tablenotemark{c}\\
$\log\mathrm{EM}_{10}$ (\cmcc)\dotfill & \nodata        			     	& $54.46 \pm 0.10$\tablenotemark{c}		& $53.42 \quad (54.08 - 53.90)\tablenotemark{d}$      	& {} & \nodata  			& $54.68 \pm 0.21$\tablenotemark{c}\\
$\log\mathrm{EM}_{11}$ (\cmcc)\dotfill & \nodata        			     	& $55.00 \pm 0.13$\tablenotemark{c}		& $53.61 \quad (53.74 - 52.32)\tablenotemark{d}$      	& {} & \nodata  			& $54.12 \pm 0.21$\tablenotemark{c}\\
$\log\mathrm{EM}_{12}$ (\cmcc)\dotfill & \nodata        			     	& $54.66 \pm 0.12$\tablenotemark{c}		& $53.76 \quad (52.52 - 52.72)\tablenotemark{d}$      	& {} & \nodata  			& $52.85 \pm 0.25$\tablenotemark{c}\\
$\log\mathrm{EM}_{13}$ (\cmcc)\dotfill & \nodata        			     	& $53.29 \pm 0.18$\tablenotemark{c}		& $53.91 \quad (52.92 - 53.07)\tablenotemark{d}$      	& {} & \nodata  			& $54.11 \pm 0.34$\tablenotemark{c}\\
$\log\mathrm{EM}_{14}$ (\cmcc)\dotfill    & \nodata				        & \nodata					& $54.12 \quad (53.14 - 53.05)\tablenotemark{d}$      	& {} & \nodata  			& \nodata			   \\
$\log\mathrm{EM}_{15}$ (\cmcc)\dotfill     & \nodata				        & \nodata					& $54.30 \quad (53.10 - 53.23)\tablenotemark{d}$      	& {} & \nodata  			& \nodata			   \\
$\log\mathrm{EM}_{16}$ (\cmcc)\dotfill    & \nodata				        & \nodata					& $54.40 \quad (53.42 - 53.76)\tablenotemark{d}$      	& {} & \nodata  			& \nodata			   \\
$\log\mathrm{EM}_{17}$ (\cmcc)\dotfill    & \nodata				        & \nodata					& $54.45 \quad (53.83 - 53.93)\tablenotemark{d}$      	& {} & \nodata  			& \nodata			   \\
$\log\mathrm{EM}_{18}$ (\cmcc)\dotfill    & \nodata				        & \nodata					& $54.41 \quad (54.15 - 54.41)\tablenotemark{d}$      	& {} & \nodata  			& \nodata			   \\
$\log\mathrm{EM}_{19}$ (\cmcc)\dotfill    & \nodata				        & \nodata					& $54.33 \quad (54.53 - 54.51)\tablenotemark{d}$      	& {} & \nodata  			& \nodata			   \\
$\log\mathrm{EM}_{20}$ (\cmcc)\dotfill    & \nodata				        & \nodata					& $54.23 \quad (54.47 - 54.40)\tablenotemark{d}$      	& {} & \nodata  			& \nodata			   \\
$\log\mathrm{EM}_{21}$ (\cmcc)\dotfill    & \nodata				        & \nodata					& $54.08 \quad (54.31 - 54.16)\tablenotemark{d}$	& {} & \nodata  			& \nodata			   \\
Statistics $/$ d.o.f\dotfill           & $5703/5226$        			     	& $5784/5219$					& $5947\tablenotemark{d}/$\nodata			& {} & $1501/1260$			& $1514/1260$ 		     
\enddata																	
\tablenotetext{a}{M1 and M2 obtained absolute abundances, [A/H], however we give in this Table [A/Fe] and [Fe/H] since				
ratios are more stable. The absolute abundances and their uncertainties
can be calculated back, $\mathrm{[A/H] = [A/Fe] + [Fe/H]}$ and $\mathrm{\Delta [A/H]_+ = \Delta [A/Fe]_+ - \Delta [Fe/H]_-}$, 
$\mathrm{\Delta [A/H]_- = \Delta [A/Fe]_- - \Delta [Fe/H]_+}$, respectively, where the $+$ and
$-$ subscripts refer to the positive and negative error bars, respectively.}
\tablenotetext{b}{The colon indicates that the value was fixed, and corresponds to the middle of the temperature bin.}
\tablenotetext{c}{Obtained from the coefficients $a_1,\dots,a_7$.}
\tablenotetext{d}{EMD$(T)$ for the best-fit line fluxes and its range (see text for details). As a quality measure
of method 3, we give the $C$ statistics after convolution of the model
through the \chan\  response matrices.}
\tablecomments{i) We give the EM per bin for methods 2 and 3 as defined in Eq.~\ref{eq:emd}. ii) Coronal abundances were compared to the solar photospheric abundances from \citet{grevesse98}.} 
\end{deluxetable}

\begin{deluxetable}{lrllcll}
\tablecolumns{7}
\tablewidth{0pc}
\tablecaption{\chan\ HETGS Fluxes for Lines with Excess Width.\label{tab:excesswidth}}
\tablehead{
\colhead{Ion Transition} & \colhead{$\lambda$\tablenotemark{a}} & \colhead{Flux\tablenotemark{b}} & \colhead{68\% Conf.
range\tablenotemark{c}} & \colhead{Width\tablenotemark{d}} & \colhead{68\% Conf. range\tablenotemark{c}} & \colhead{Doppler width\tablenotemark{e}}\\
\colhead{\makebox[2.0cm][c]{Line}} & \colhead{(\AA)}           & \multicolumn{2}{c}{($\mathrm{10^{-6}\; ph~cm^{-2}~s^{-1}}$)}
 & \multicolumn{2}{c}{(m\AA)} & \colhead{(m\AA)}}
\startdata
\ion{Si}{14}\dotfill & $6.18$  & $18.3$ &  $(16.5 - 20.2)$	& $4.4$ & $(3.2 - 5.4)$ & $2.2$\\
\ion{Mg}{12}\dotfill & $8.42$  & $16.0$ &  $(14.4 - 17.7)$	& $5.2$ & $(3.1 - 7.4)$ & $3.2$\\
\ion{Ne}{10}\dotfill & $12.13$ & $137$ &  $(130 - 143)$		& $6.2$ & $(5.0 - 7.1)$ & $5.2$\\
\ion{O}{8}\dotfill   & $18.97$ & $131$ &  $(112 - 153)$		& $6.3$ & $(2.7 - 9.1)$ & $9.1$\\
\ion{N}{7}\dotfill   & $24.78$ & $8.0$ &  $(6.7 - 9.5)$		& $4.3$ & $(0 - 14.1)$  & $12.7$
\enddata
\tablenotetext{a}{Wavelengths from APEC 1.3.1 database.}
\tablenotetext{b}{Corrected for photoelectric absorption ($\log N_\mathrm{H} \sim 20.87$~\cmsq).}
\tablenotetext{c}{Obtained with $\Delta C = 1$.}
\tablenotetext{d}{Gaussian width in excess of the instrumental profile.}
\tablenotetext{e}{Doppler broadening for an isothermal plasma of $40$~MK.}
\end{deluxetable}

\begin{deluxetable}{lrll}
\tablecolumns{4}
\tablewidth{0pc}
\tablecaption{\chan\ HETGS Fluxes Used For Density\- and Opacity-Sensitive Line
Ratios.\label{tab:chanfluxes}}
\tablehead{
\colhead{Ion Transition} & \colhead{$\lambda$\tablenotemark{a}} & \colhead{Flux\tablenotemark{b}} & \colhead{68\% Conf.
range\tablenotemark{c}}\\
\colhead{\makebox[2.0cm][c]{Line}} & \colhead{(\AA)}           & \multicolumn{2}{c}{($\mathrm{10^{-6}\; ph~cm^{-2}~s^{-1}}$)}}
\startdata
\ion{Si}{13}\dotfill & $6.65$ & $8.0$ &  $(6.7 - 9.5)$\\
\ion{Si}{13}\dotfill & $6.69$ & $0.9$ &  $(0 - 2.0)$\\
\ion{Si}{13}\dotfill & $6.74$ & $6.8$ &  $(5.5 - 8.2)$\\

\ion{Mg}{11}\dotfill & $9.17$  & $5.5$ &  $(4.1 - 7.0)$\\
\ion{Mg}{11}\dotfill & $9.23$  & $1.8$ &  $(0.6 - 3.1)$\\
\ion{Mg}{11}\dotfill & $9.31$  & $1.1$ &  $(0 - 2.5)$\\

\ion{Ne}{9}\dotfill  & $13.45$ & $12.9$ &  $(9.7 - 16.5)$\\
\ion{Ne}{9}\dotfill  & $13.55$ & $0$	&  $(0 - 1.8$\\
\ion{Ne}{9}\dotfill  & $13.70$ & $6.1$ &  $(3.4 - 9.2)$\\

\ion{O}{7}\tablenotemark{d}\dotfill   & $21.60$ & $24.4$ &  $(12.5 - 39.8)$\\
\ion{O}{7}\tablenotemark{d}\dotfill   & $21.80$ & $24.7$ &  $(12.3 - 40.9)$\\
\ion{O}{7}\tablenotemark{d}\dotfill   & $22.10$ & $5.0$ &     $(0 - 16.0)$\\

\ion{Ne}{10}         & $10.24$ & $16.4$ & $(14.3 - 18.6)$\\
\ion{Ne}{10}         & $12.13$& $137$  & $(130 - 143)$\\

\ion{O}{8}           & $16.00$ & $29.2$ & $(23.8 - 35.1)$\\
\ion{O}{8}           & $18.97$ & $131$  & $(112 - 153)$\\

\ion{Fe}{17}\dotfill & $15.01$ & $19.0$ & $(14.9 - 23.6)$\\
\ion{Fe}{17}\dotfill & $15.26$ & $7.2$ & $(4.0 - 11.0)$\\
\ion{Fe}{17}\dotfill & $16.76$ & $11.8$ & $(7.0 - 17.3)$

\enddata
\tablenotetext{a}{Wavelengths from APEC 1.3.1 database.}
\tablenotetext{b}{Corrected for photoelectric absorption ($\log N_\mathrm{H} \sim 20.87$~\cmsq).}
\tablenotetext{c}{Obtained with $\Delta C = 1$.}
\tablenotetext{d}{Very weak signal. Essentially $\sim 3$ counts are detected above the continuum
	for the $r$ and $i$ lines.}
\end{deluxetable}

\pagebreak
\newpage

\renewcommand{\thefigure}{\arabic{figure}\alph{subfigure}}
\setcounter{subfigure}{1}
\begin{figure}
\centering
\includegraphics[width=\textwidth]{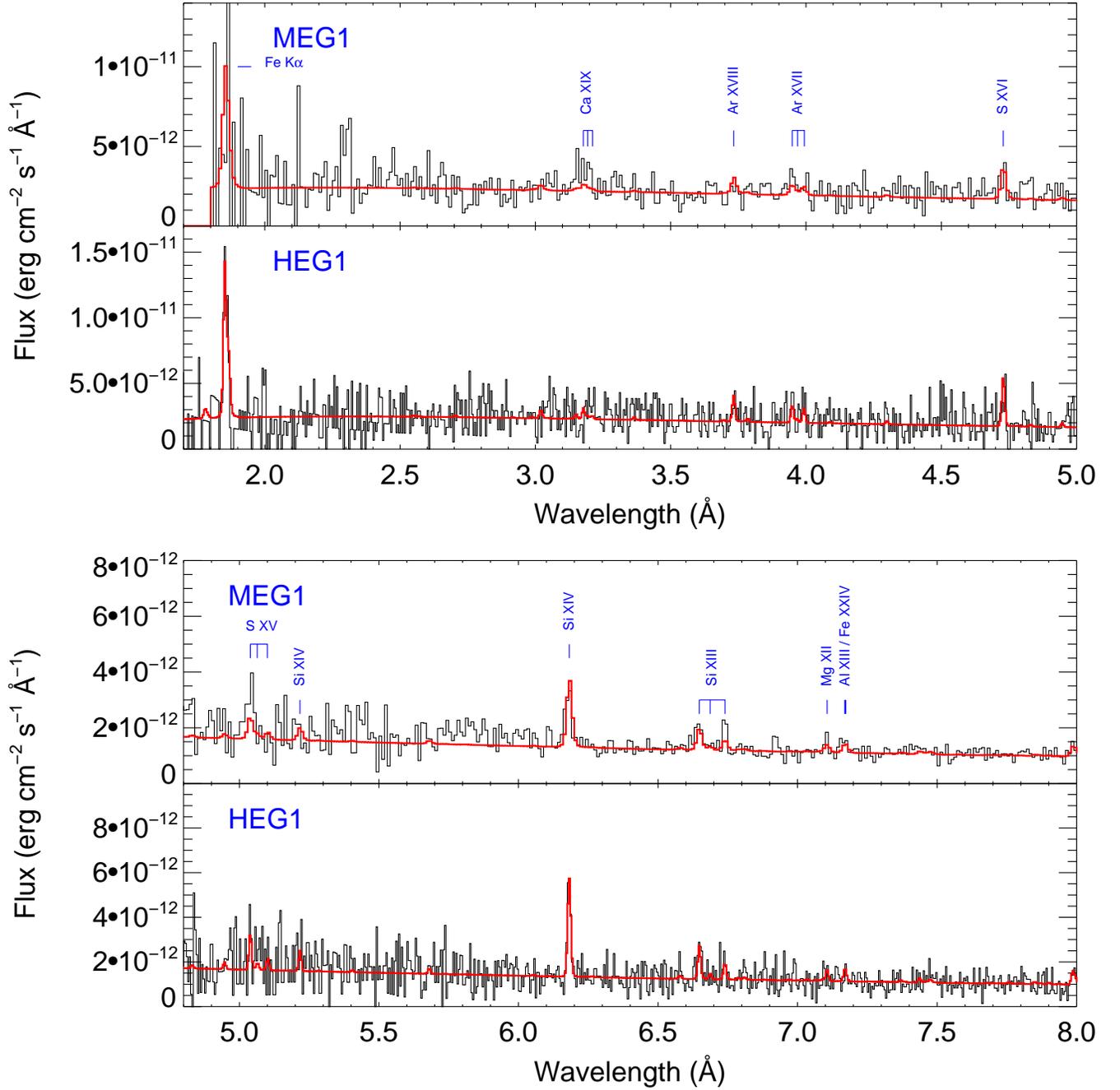}
\caption{($a$) \chan\  MEG1 and HEG1 spectra compared with the best-fit 3-$T$ model 
(red line) found in Tab.~\ref{tab:meth}. ($b$) Same as
($a$), but for $8-14$~\AA. ($c$) Same as ($a$) but for $14-26$~\AA. The $17-26$~\AA\ 
panels show the MEG data only.\label{fig:chandra_bfit_meth1a}}
\end{figure}
\newpage

\addtocounter{figure}{-1}
\addtocounter{subfigure}{1}
\begin{figure}
\centering
\includegraphics[width=\textwidth]{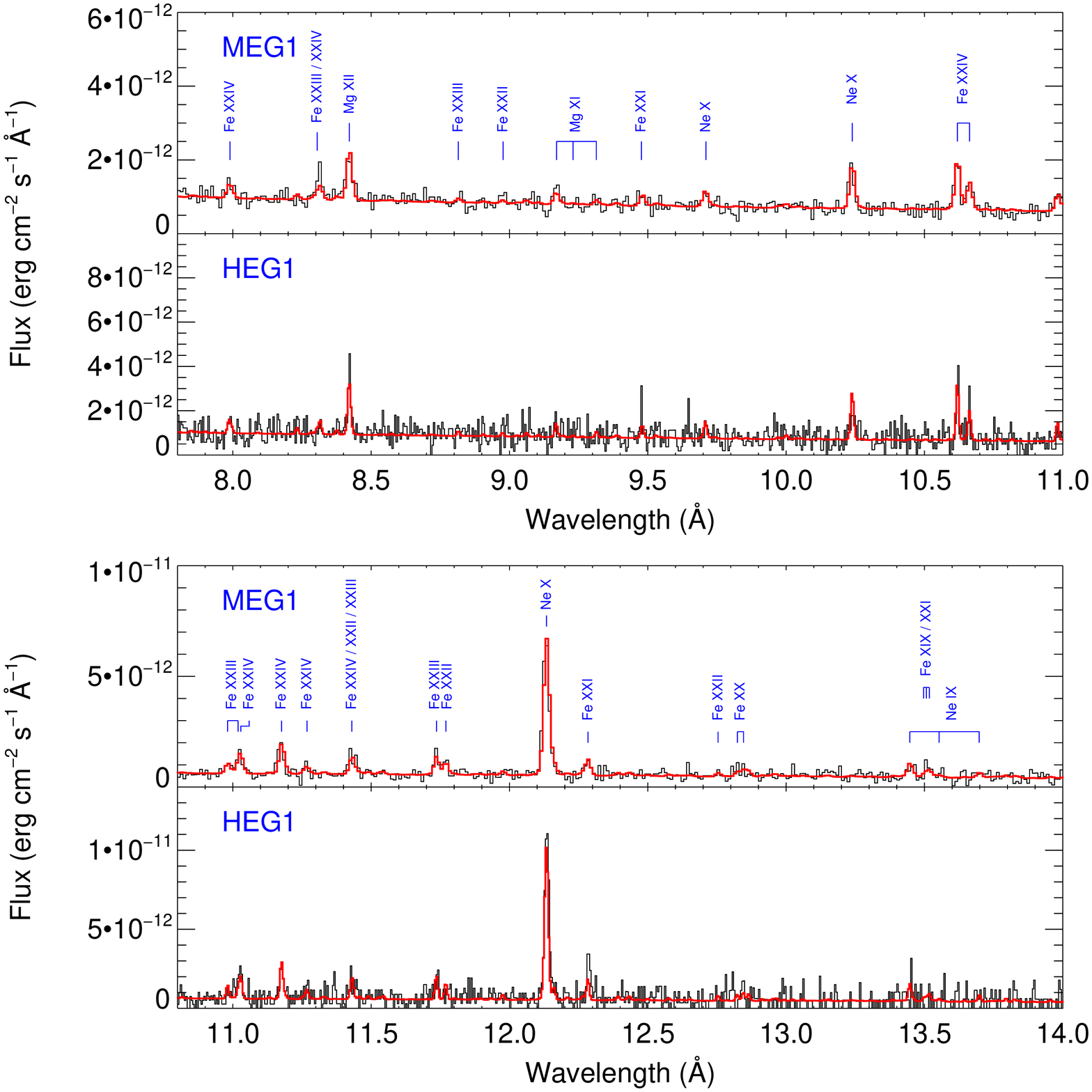}
\caption{\label{fig:chandra_bfit_meth1b}}
\end{figure}
\newpage

\addtocounter{figure}{-1}
\addtocounter{subfigure}{1}
\begin{figure}
\centering
\includegraphics[width=\textwidth]{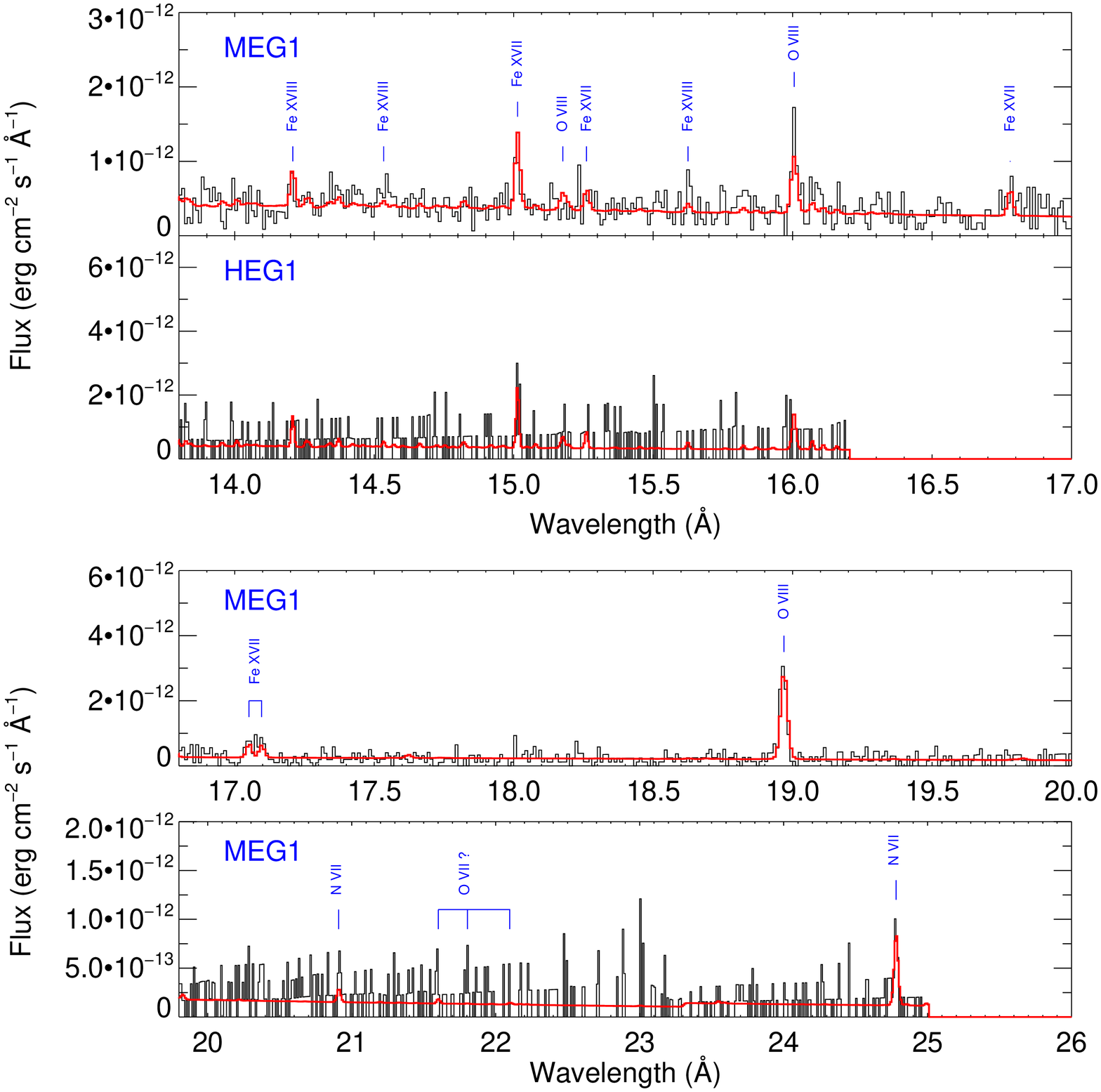}
\caption{\label{fig:chandra_bfit_meth1c}}
\end{figure}
\newpage

\renewcommand{\thefigure}{\arabic{figure}}
\begin{figure}
\centering
\includegraphics[width=\textwidth]{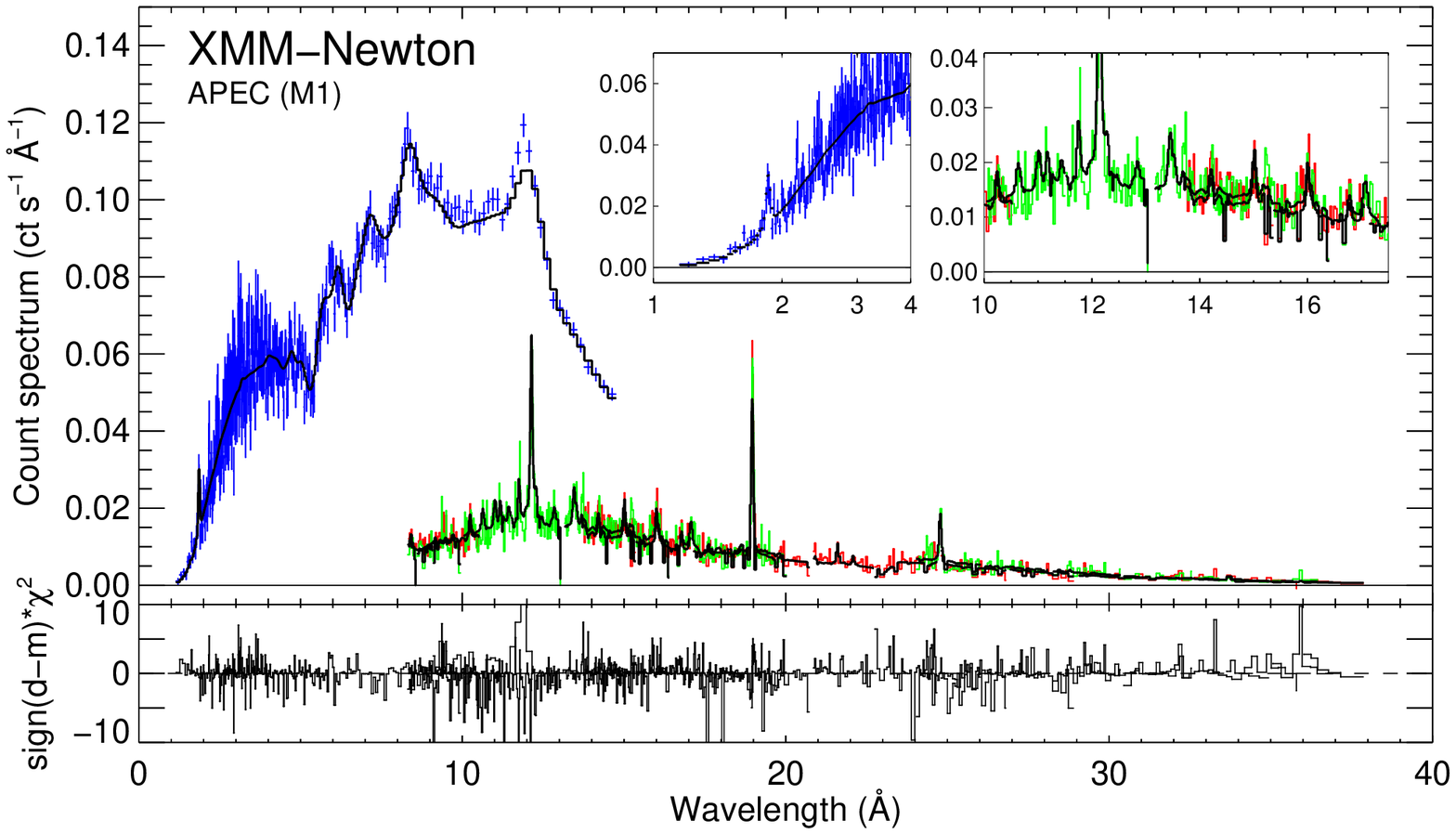}
\caption{\xmm\ MOS1, RGS1, and RGS2 spectra of YY Men with the best-fit 4-$T$ 
model of method 1 overlaid as a thick black line. Error bars are plotted for MOS only, for 
clarity. Contributions to the $\chi^2$ (multiplied by sign[data - model]) are 
also plotted in the lower panel. Zoom-ins from 1~\AA\  to 4~\AA\  (logarithmic wavelength scale)
and from 10~\AA\  to 17.5~\AA\  (linear wavelength scale) show the quality of the fits in the 
region where highly ionized Fe lines are emitted.\label{fig:xmmspec}}
\end{figure}
\newpage

\begin{figure}
\centering
\includegraphics[width=\textwidth]{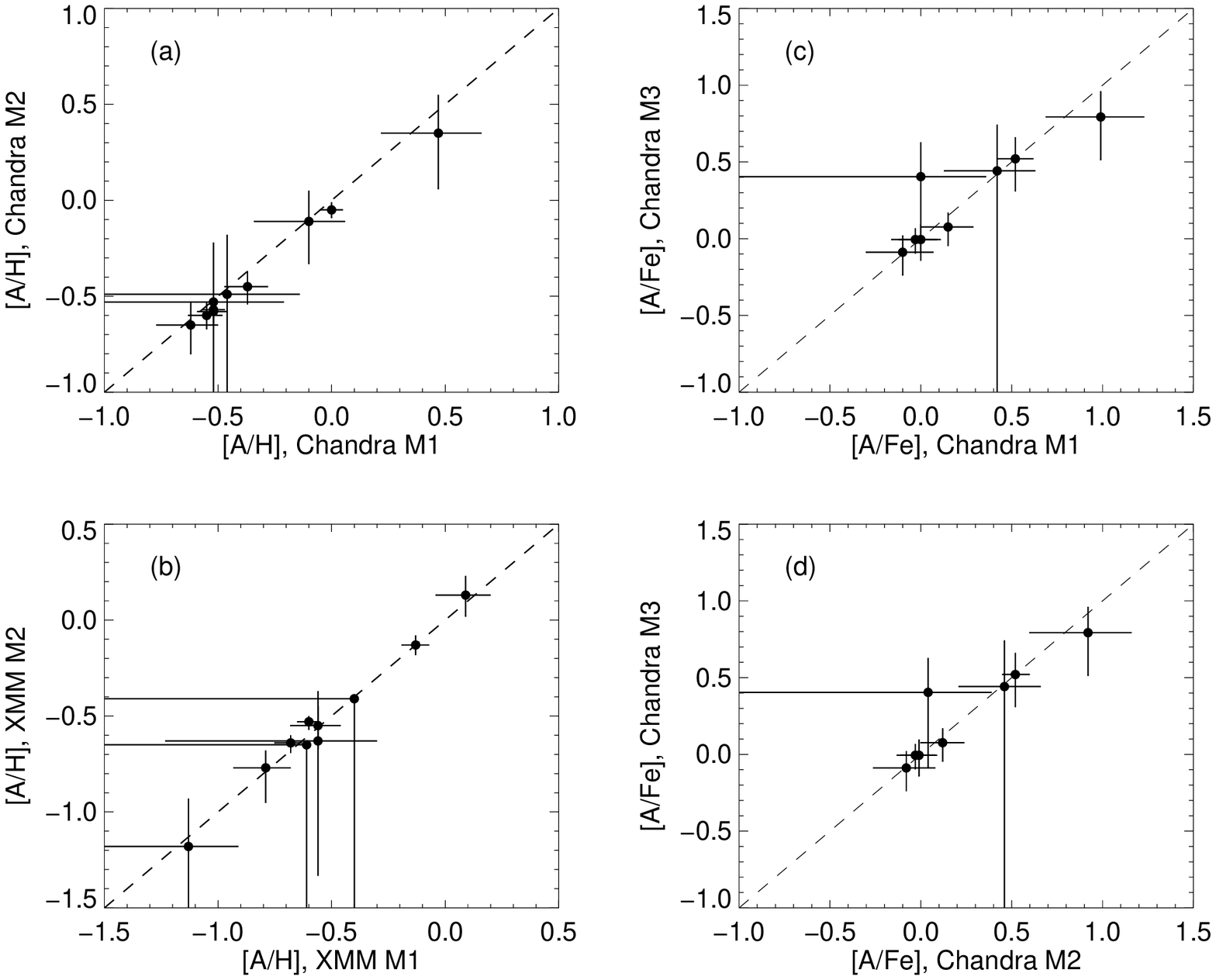}
\caption{{\it Left:} Comparison between absolute abundances derived from methods 1 and 2
with \chan\  (top panel) and \xmm\  (bottom panel) data. The dashed line
represents a 1:1 correlation.  Absolute abundances from method 2 are
slightly lower with respect to those obtained from method 1 for \chan\  data (but
not for \xmm\  data).
{\it Right:} Comparison between abundance ratios derived 
from method 3 and methods 1 (top) and 2 (bottom) with \chan\  data. 
Excellent agreement is generally found.
\label{fig:ab_m12_chan123}}
\end{figure}
\newpage

\begin{figure}
\centering
\includegraphics[width=\textwidth]{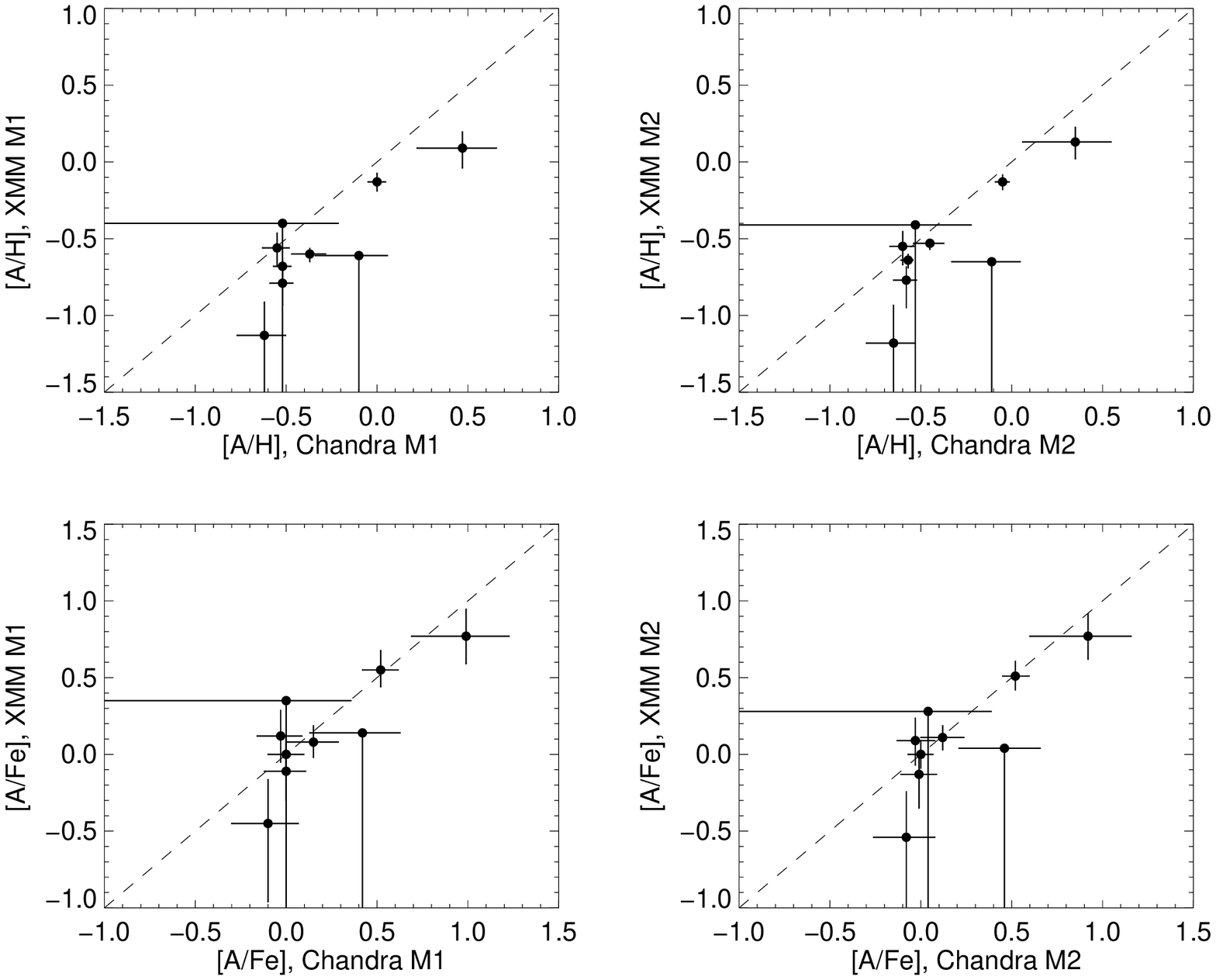}
\caption{Comparison between abundances derived from methods 1 (left
panels) and 2 (right panels) with \xmm\  and \chan\  data. The dashed line
represents a 1:1 correlation. The upper panels show the absolute abundances (relative to
H), and the lower panels show the abundances relative to Fe ([A/Fe] = [A/H] $-$
[Fe/H]). All abundances are relative to the solar photospheric composition
\citep{grevesse98}.
\label{fig:ab_xmmchan}}
\end{figure}
\newpage

\begin{figure}
\centering
\includegraphics[width=\textwidth]{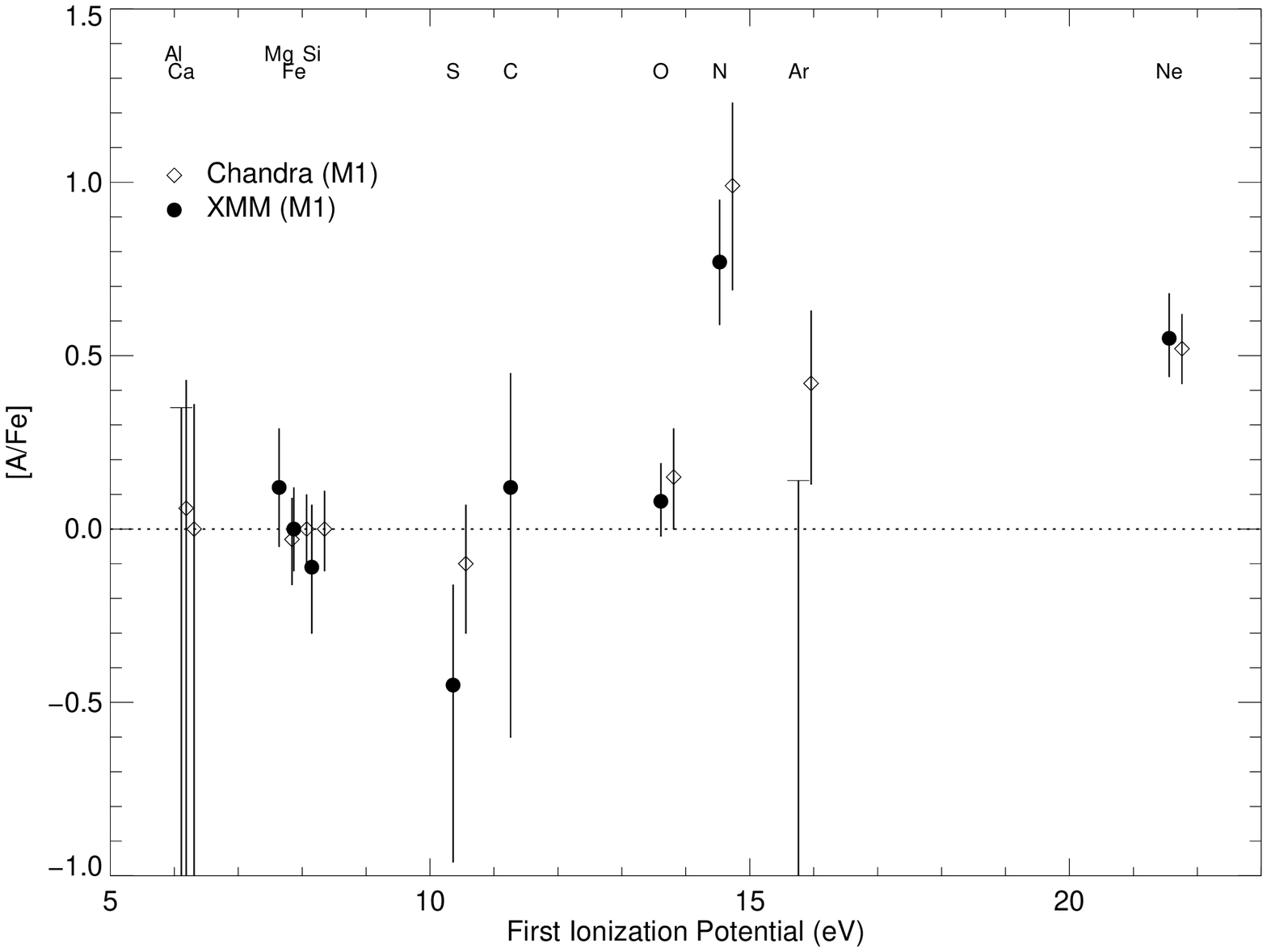}
\caption{Abundance ratios, [A/Fe], in YY Men's corona as a function of the FIP.
For clarity, we selected abundances from method 1, and shifted the \chan\  data
points by $+0.2$~eV. The abundance ratios are relative to the solar photospheric 
composition \citep{grevesse98}. 
 The solar ratio ($\mathrm{[A/Fe] = 0}$) is 
indicated by a dotted line.\label{fig:ab_fip}}
\end{figure}
\newpage

\begin{figure}
\centering
\includegraphics[width=0.9\textwidth]{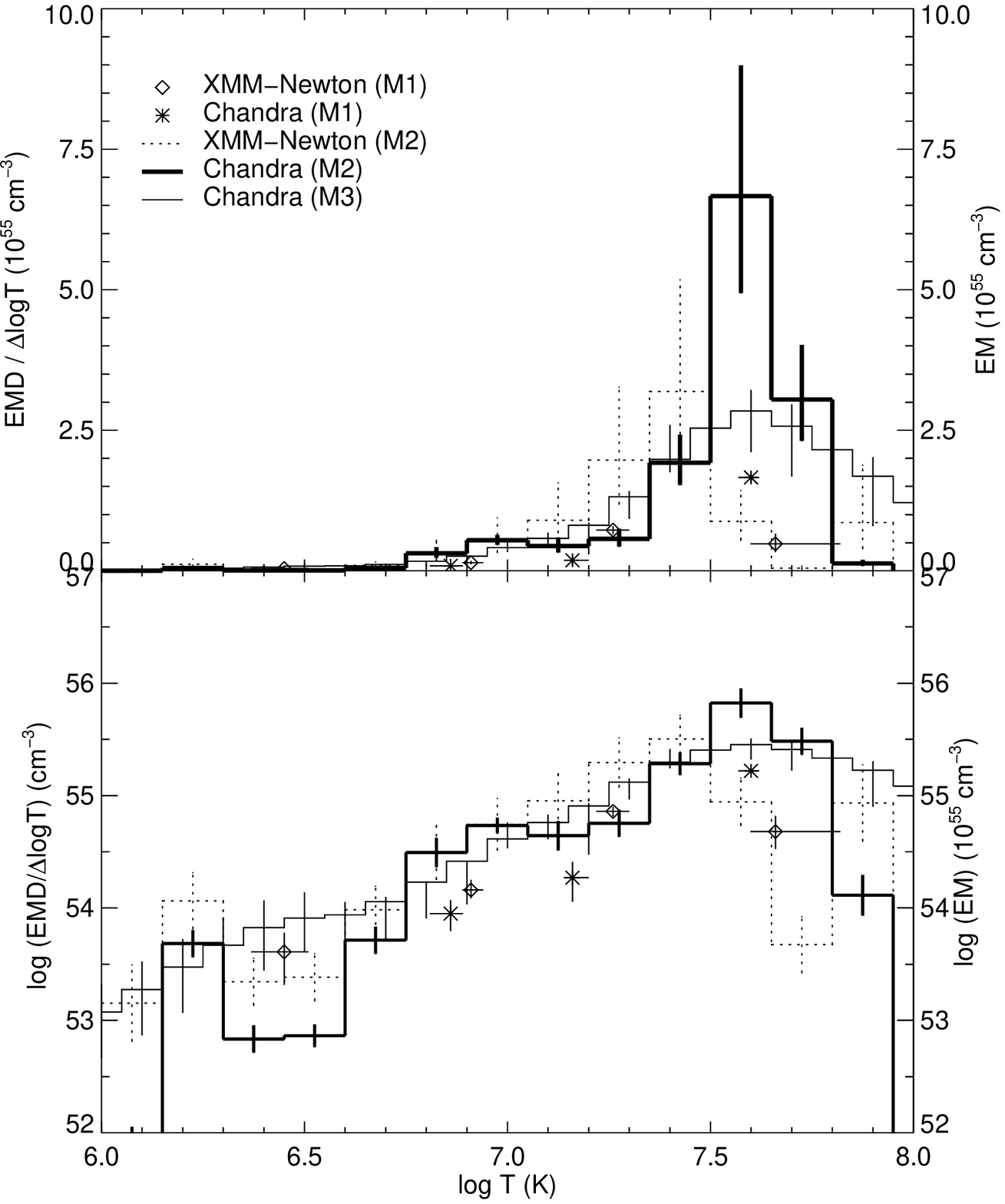}
\caption{Emission measure distribution of YY Men obtained from the \chan\  and \xmm\  spectra for the 
various methods. To help comparison, we plot  $\mathrm{EMD}(T)/\Delta\log T$, i.e., the EMDs per bin 
(see Table~\ref{tab:meth}) were divided by the integration step ($0.15$ and $0.10$ dex for methods 2 and 3, 
respectively). For the multi-T approach (method 1), we show the EM in each component since this method has 
no bin width. The y-axis, therefore, reflects this distinction (left axis for methods 2 and 3, right axis 
for method 1). 
The top panel shows the EMD in a linear vertical scale to
emphasize the dominant very hot plasma, whereas the lower panel uses a logarithmic
vertical scale to reveal the weak, but required plasma at lower temperatures.
\label{fig:emd_norm}}
\end{figure}
\newpage

\begin{figure}
\centering
\includegraphics[width=\textwidth]{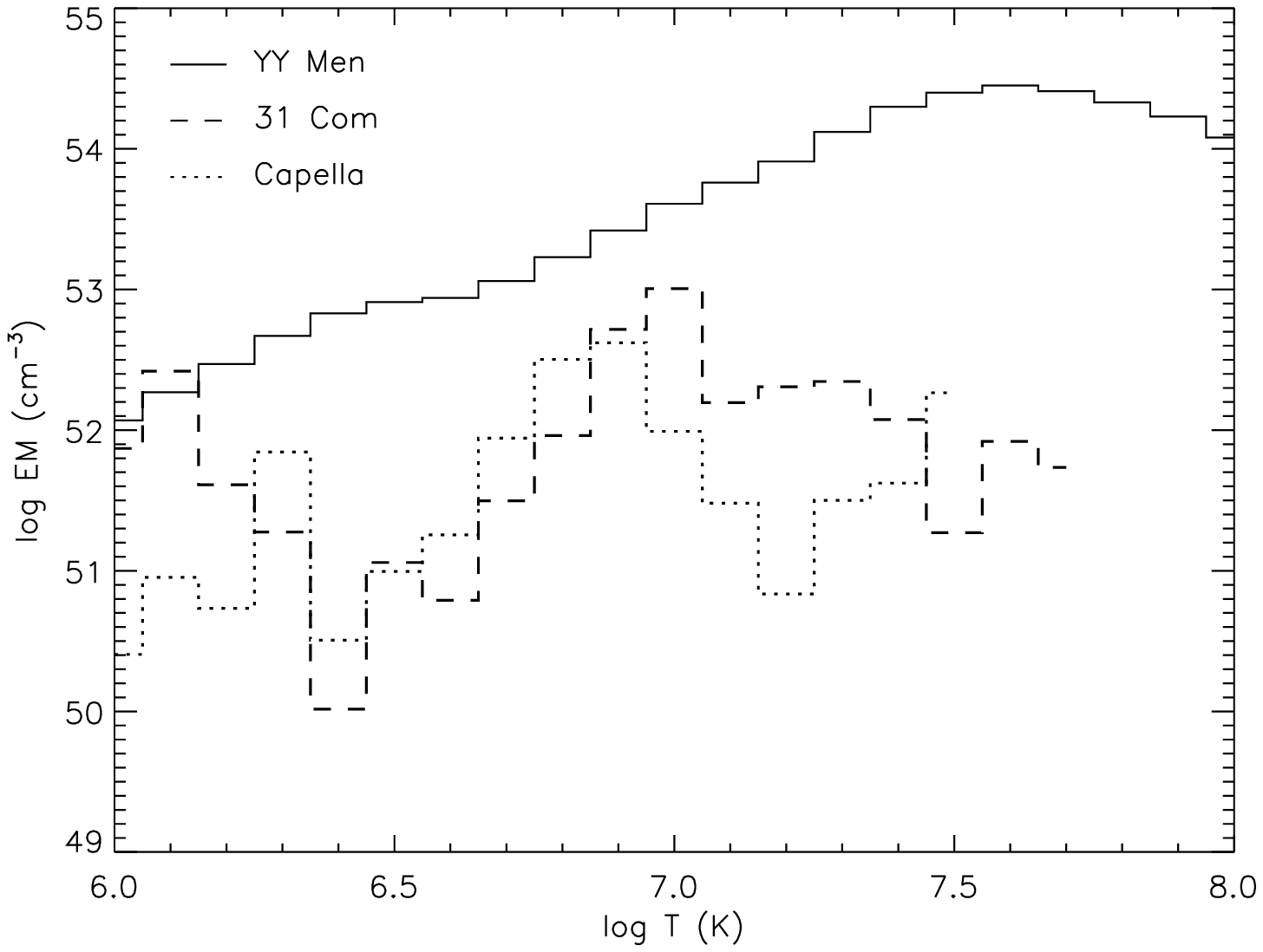}
\caption{Comparison of the EMDs obtained in this paper for
YY Men (method 3, using the definition in Eq.~\ref{eq:emd}) and those derived for the giants Capella, and 31 Com \citep{argiroffi03,scelsi04}.
Note the much flatter EMD, the higher
temperatures, and activity level of YY Men.\label{fig:emdcompare}}
\end{figure}
\newpage

\begin{figure}
\centering
\includegraphics[width=\textwidth]{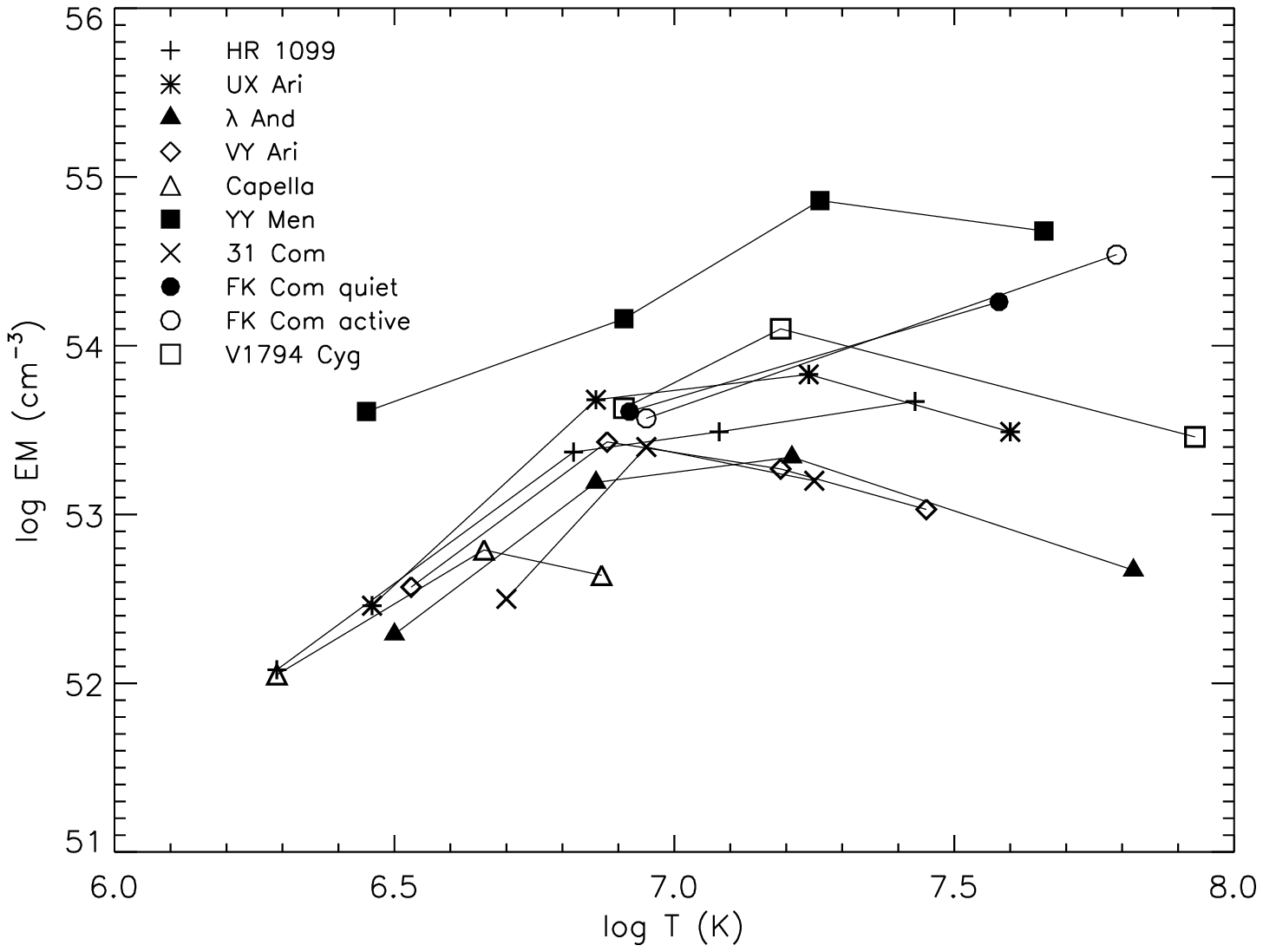}
\caption{Comparison of the results of $4$-$T$ fits of YY Men \xmm\  spectra
with those obtained by \citet{audard03} for a sample of 
RS CVn stars of different activity levels. Also shown here
is a $3$-$T$ fit of the single rapidly rotating giant 
31 Com \citep{scelsi04}. Fits for two other FK Com-type stars (V1794 Cyg and FK Com) from \citet{gondoin04}
and \citet{gondoin02} are also shown. The lines connecting the data points of each star are
an aid to the eye and have no physical meaning.\label{fig:emdcompareall}}
\end{figure}
\newpage

\begin{figure}
\centering
\includegraphics[width=0.8\textwidth]{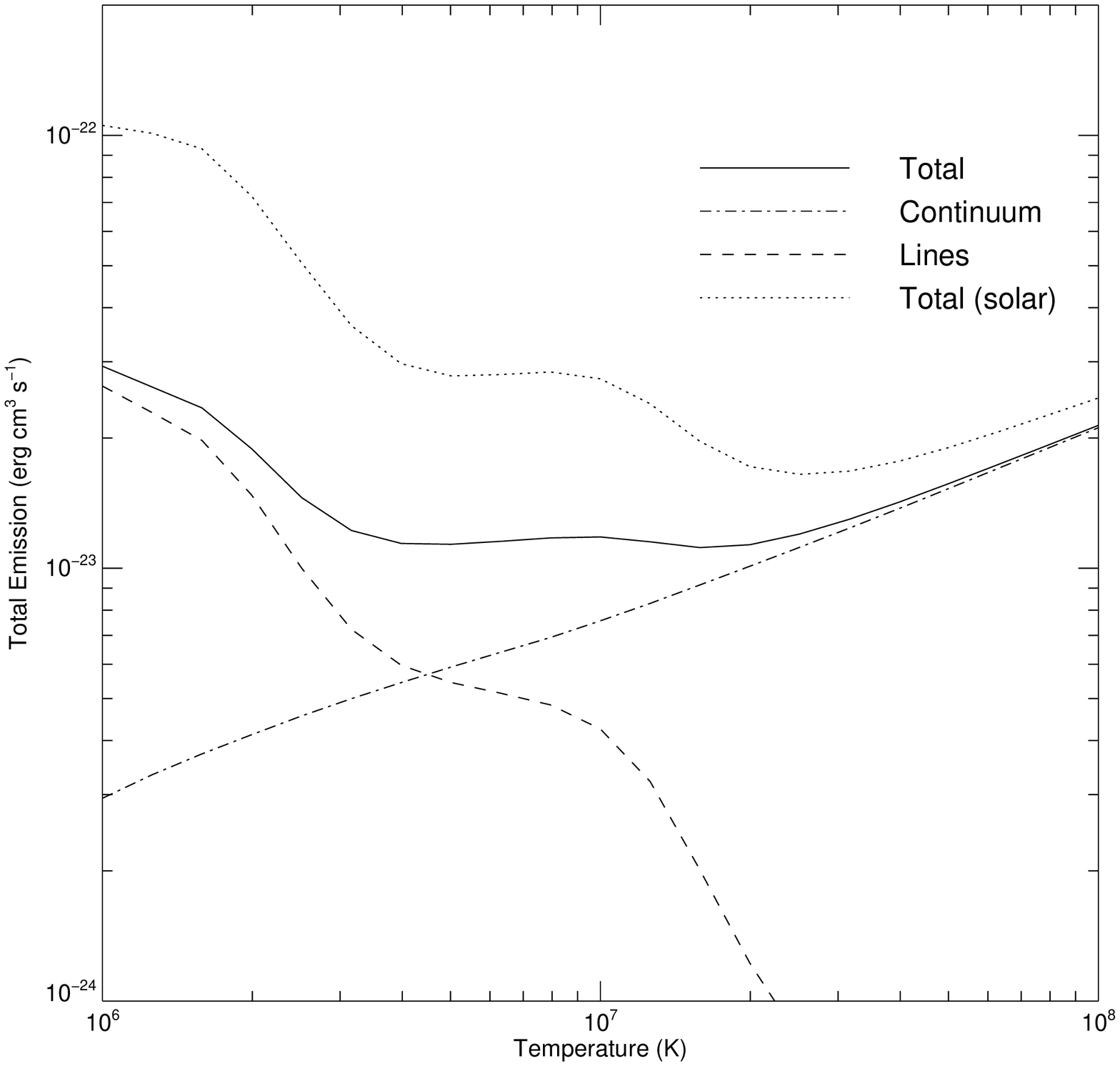}
\caption{Radiative cooling curve in units of erg~cm$^3$~s$^{-1}$ for $T=1 -
100$~MK, based on the
APEC 1.3.1 database (for photon energies from 0.01~keV to 50~keV, i.e.,
0.25~\AA\  to 1240~\AA) and for best-fit \xmm\  abundances (M1; Tab.~\ref{tab:meth}). 
The total radiative loss function (solid) is the sum of the
continuum contribution (dash-dotted) and of the line
contribution (dashed).  For comparison the total radiative loss for solar
abundances \citep{grevesse98} is shown as a dotted line.\label{fig:coolcurve}}
\end{figure}
\newpage

\begin{figure}
\centering
\includegraphics[width=0.9\textwidth]{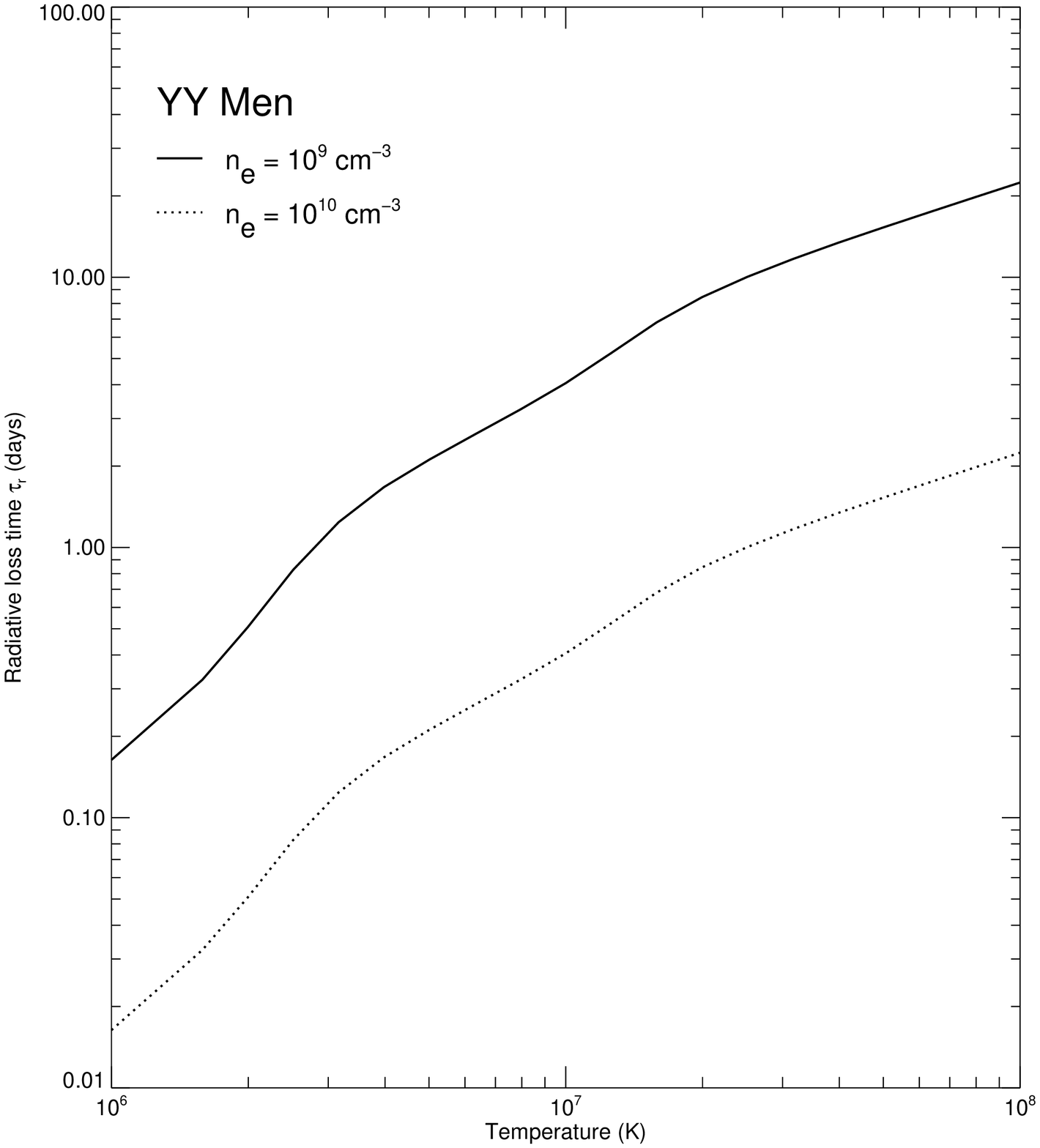}
\caption{Radiative loss time, $\tau_r(T) = 3kT/(n_\mathrm{e}\Lambda(T))$, in units of days derived 
from the radiative cooling curve, $\Lambda(T)$, in YY Men's corona as shown in Fig.~\ref{fig:coolcurve}.
Two different plasma electron densities, $n_\mathrm{e} = 10^9$ and $10^{10}$~\cmcc,  were used. As mentioned
in Sect.~\ref{sect:densities}, there is no evidence for higher densities in YY Men.
\label{fig:radcooltime}}
\end{figure}
\clearpage

\begin{figure}
\centering
\includegraphics[width=0.9\textwidth]{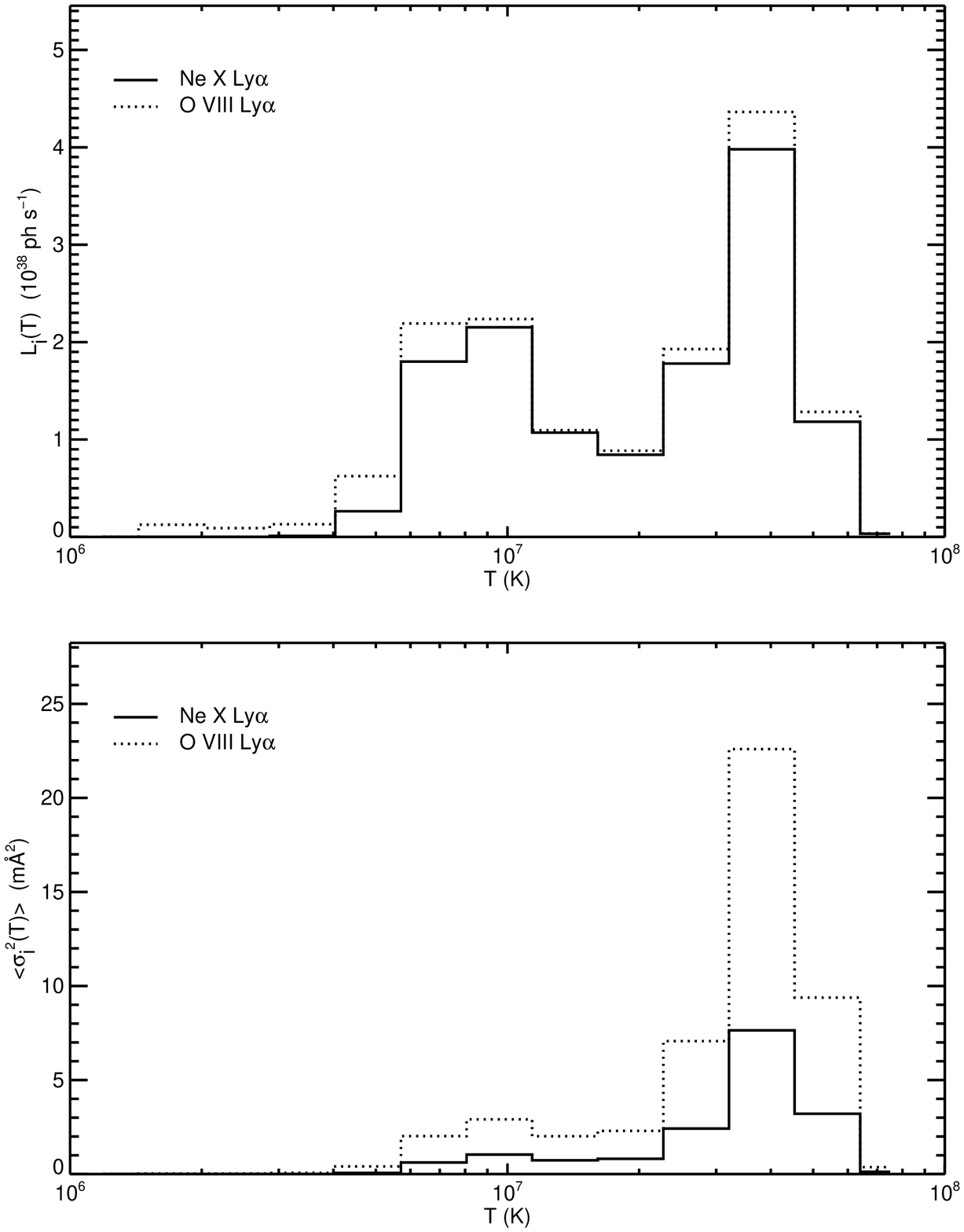}
\caption{\textit{(Top panel):} Line photon luminosity distribution for
\ion{Ne}{10} Ly$\alpha$ (solid) and \ion{O}{8} Ly$\alpha$ (dotted).
\textit{(Bottom panel):} Weighted average width distribution for the same lines (see text for details).
\label{fig:lumline}}
\end{figure}
\clearpage

\begin{figure}
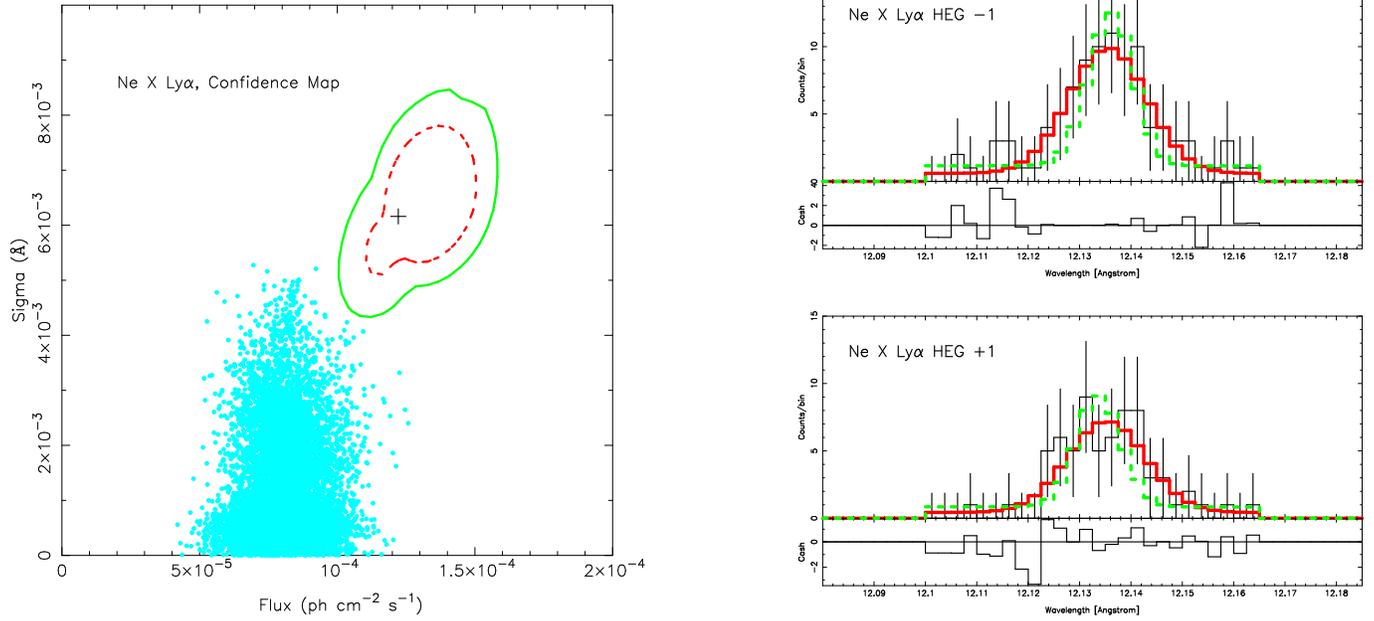

\centering
\includegraphics[width=.47\textwidth]{f13a.eps}
\hfill
\includegraphics[width=.42\textwidth]{f13b.eps}
\caption{{\it (Left)} Confidence map for the \ion{Ne}{10} Ly$\alpha$ line. Confidence
contours with $\Delta C = 1$ and $\Delta C = 2.71$ are shown in red (inner dotted contour) and green (outer solid
contour), respectively.
The loci of the best-fit parameters for 10,000 Monte-Carlo simulations of an emission line with
an instrumental profile are shown as cyan-colored dots. {\it (Right)} HEG line profile of \ion{Ne}{10} Ly$\alpha$ in the negative (top) and positive 
(bottom) order spectra. The best-fit Gaussian profile is shown as a red line (wide solid line), 
whereas the best-fit  with the instrumental profile is shown as a green line (narrow dashed line).
Residuals from the fit with a Gaussian profile are shown in sub-panels.\label{fig:NeXLya}}
\end{figure}
\clearpage

\begin{figure}
\centering
\includegraphics[width=.47\textwidth]{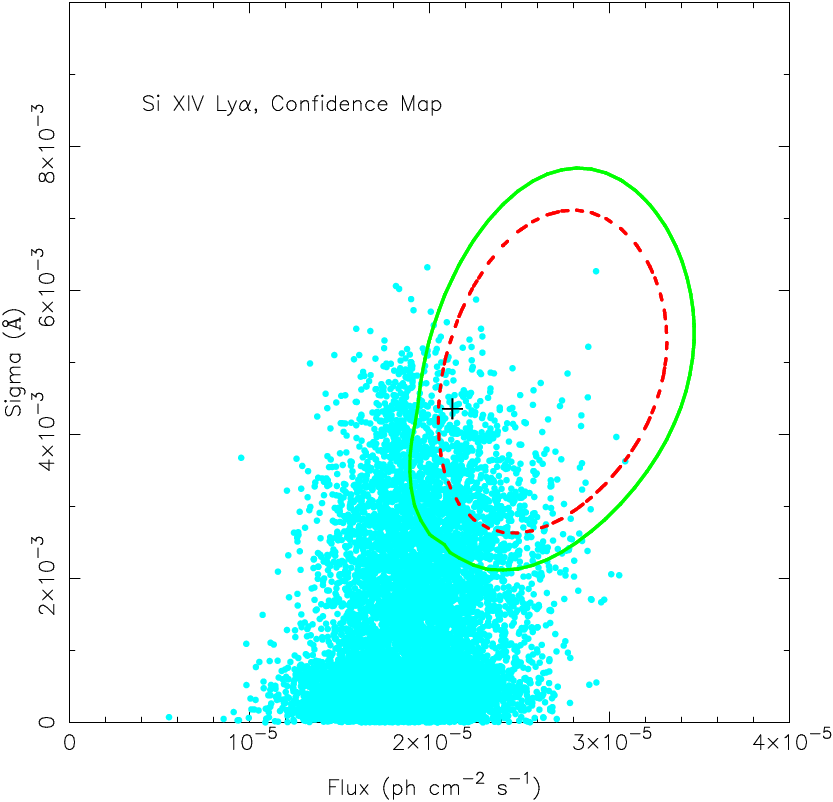}
\hfill
\includegraphics[width=.42\textwidth]{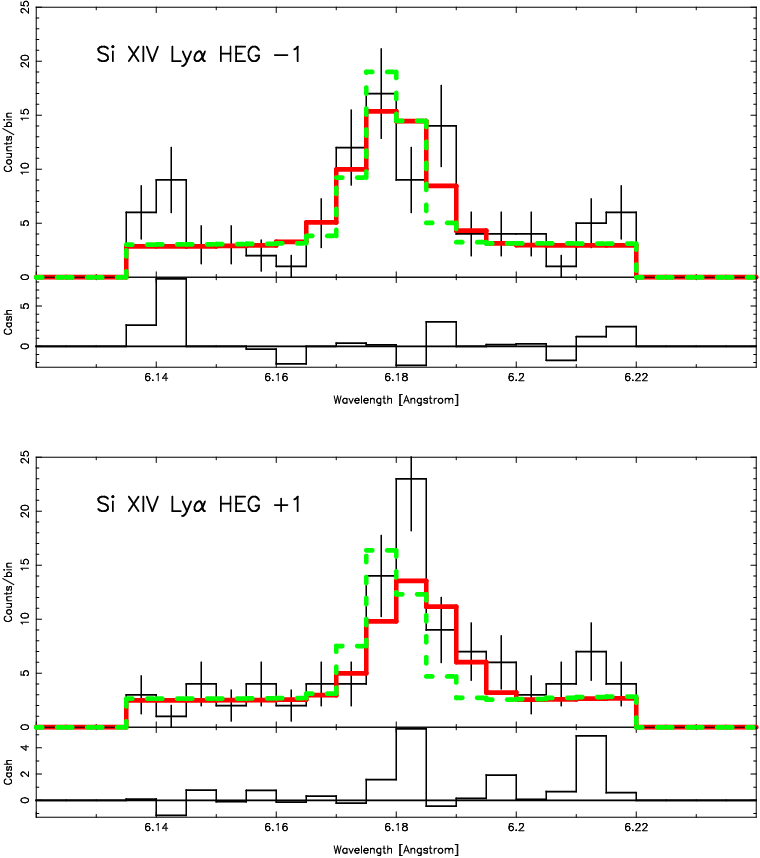}
\caption{Similar to Figure~\ref{fig:NeXLya} but for \ion{Si}{14} Ly$\alpha$.\label{fig:SiXIVLya}}
\end{figure}

\end{document}